%% file: main.tex
\documentclass[a4paper,11pt]{article}
\pdfoutput=1
\usepackage{jcappub}
\usepackage[table,svgnames,dvipsnames]{xcolor}

\usepackage{hyperref}
\usepackage{orcidlink}
\usepackage{hanging}
\usepackage{aas_macros}
\usepackage{amsmath}
\usepackage{amssymb}
\usepackage[utf8]{inputenc}
\usepackage[a4paper, left=2.5cm, right=2.5cm, bottom=2.5cm, top=2.5cm]{geometry}
\usepackage{comment}
\usepackage{graphicx}
\usepackage{bbold}
\usepackage[normalem]{ulem}
\usepackage{gensymb}
\graphicspath{ {figures/} }

\newcommand{\hMpc}{\ h\,\text{Mpc}^{-1}}
\newcommand{\hhhMpc}{\ h^3\,\text{Mpc}^{-3}}
\newcommand{\Mpch}{\ h^{-1}\text{Mpc}}
\newcommand{\Gpch}{\ h^{-1}\text{Gpc}}
\newcommand{\sigYone}{$\sigma^\mathrm{DR1}_{\mathrm{stat}}$}
\newcommand{\LCDM}{$\Lambda$CDM}
\newcommand{\Abacus}{\texttt{AbacusSummit}\ }
\newcommand{\Velocileptors}{\texttt{velocileptors}\ }

\renewcommand*\vec[1]{\ensuremath{\boldsymbol{#1}}}
\newcommand*\tens[1]{\ensuremath{\mathsf{#1}}}

\usepackage[nameinlink,noabbrev]{cleveref}
\crefname{equation}{Eq.}{Eqs.}
\crefname{section}{Section}{Sections}
\crefname{figure}{Figure}{Figures}
\crefname{table}{Table}{Tables}
\crefname{appendix}{Appendix}{Appendices}
\Crefname{figure}{Figure}{Figures}
\Crefname{equation}{Equation}{Equations}
\Crefname{section}{Section}{Sections}
\Crefname{table}{Table}{Tables}

\title{Exploring HOD-dependent systematics for the DESI 2024 Full-Shape galaxy clustering analysis}

\input{HOD_author_list}

\abstract{
We analyse the robustness of the DESI 2024 cosmological inference from the full shape of the galaxy power spectrum to uncertainties in the Halo Occupation Distribution (HOD) model of the galaxy-halo connection and the choice of priors on nuisance parameters. We assess variations in the recovered cosmological parameters across a range of mocks populated with different HOD models and find that shifts are often greater than 20\% of the expected statistical uncertainties from the DESI data. We encapsulate the effect of such shifts in terms of a systematic covariance term, $\tens{C}_{\rm HOD}$, and an additional diagonal contribution quantifying the impact of our choice of nuisance parameter priors on the ability of the effective field theory (EFT) model to correctly recover the cosmological parameters of the simulations. These two covariance contributions are designed to be added to the usual covariance term, $\tens{C}_{\rm stat}$, describing the statistical uncertainty in the power spectrum measurement, in order to fairly represent these sources of systematic uncertainty. This novel approach should be more general and robust to the choice of model or additional external datasets used in cosmological fits than the alternative approach of adding systematic uncertainties to the recovered marginalised parameter posteriors. We compare the approaches within the context of a fixed \LCDM\ model and demonstrate that our method gives conservative estimates of the systematic uncertainty that nevertheless have little impact on the final posteriors obtained from DESI data.
}

\begin{document}
\maketitle
\flushbottom

\section{Introduction}
\label{sec:intro}

% Background and motivation
With the advent of the Dark Energy Spectroscopic Instrument (DESI) spectroscopic galaxy survey \cite{Snowmass2013.Levi, DESI2016a.Science, DESI2022.KP1.Instr}, we are able to significantly improve constraints on our cosmological model. By measuring the redshifts of over 50 million galaxies spanning a 5-year survey, facilitated by a robotically-controlled fibre system \cite{DESI2016b.Instr, FocalPlane.Silber.2023, Corrector.Miller.2023}, DESI is mapping the cosmic web of structure with unprecedented accuracy. The effects of both primordial and late-time physics, such as gravity and cosmic expansion, are imprinted on the large-scale distribution of galaxies. By probing this distribution out to redshift $z>2$, DESI is able to extract a wealth of cosmological information and provide a view into the history of the Universe over the past 10 billion years. The survey has already made outstanding progress towards its science goals having completed survey validation \cite{DESI2023a.KP1.SV}, produced an early public release of the data \cite{DESI2023b.KP1.EDR} and published its first cosmological results analysing the Baryon Acoustic Oscillation (BAO) feature \citep{DESI2024.III.KP4, DESI2024.IV.KP6, DESI2024.VI.KP7A}. DESI targets five classes of tracer: the low-redshift Bright Galaxy Survey (BGS), luminous red galaxies (LRG), emission line galaxies (ELG), quasars (QSO) and the Lyman-$\alpha$ forest. The survey operations and data reduction pipeline are detailed in \cite{SurveyOps.Schlafly.2023} and \cite{Spectro.Pipeline.Guy.2023}, respectively, while the tracer samples, creation of the large-scale structure catalogues and 2-point clustering measurements are described in \cite{DESI2024.II.KP3}. 

The method of compressing the observed galaxy field into 2-point summary statistics is well established and allows the majority of available information to be recovered. The behaviour of these compressed statistics is well understood on large scales and is sensitive to the energy content and expansion history of the Universe. Cosmological processes leave their signature on the 2-point statistics as two main features that can be probed by spectroscopic galaxy surveys: BAO \cite{Blake2003, Seo2003} and Redshift Space Distortions  (RSD; \cite{Kaiser1987}). The BAO analysis marginalises over broadband information, extracting only the BAO feature to provide a robust `standard ruler' measurement. However, additional cosmological information is contained within the shape of these statistics beyond the BAO scale. Analysis of the full shape of the Fourier-space galaxy power spectrum directly probes the matter distribution through the RSD effect but consequently requires accurate marginalisation over halo-scale physics. This paper explores the robustness of our power spectrum models to small-scale effects in support of the DESI 2024 Full-Shape galaxy clustering analysis \cite{DESI2024.V.KP5, DESI2024.VII.KP7B}. The method presented in this work can be generalised to also be applicable to the Full-Shape analysis performed in configuration-space \cite{KP5s5-Ramirez}. The DESI 2024 BAO and Full-Shape analyses, in combination with a measurement to constrain local primordial non-Gaussianity \cite{ChaussidonY1fnl}, mark the culmination of effort to shed light on the cosmological model with the first year of DESI data contained in Data Release 1 (DR1; \cite{DESI2024.I.DR1}).

% Outline the problem
On large scales, the galaxy power spectrum can be described by linear theory but the abundance of modes at smaller scales provides incentive for more complex modelling. As the Universe evolves, the initially Gaussian dark matter (DM) field undergoes non-linear evolution due to gravity, eventually clustering to form DM halos. These peaks in the density field lay the foundations for galaxy formation, although the intricacies of this process remain unclear \citep{Wechsler2018}. On large scales ($k< 0.1\hMpc$), a single linear parameter is sufficient to describe the bias of galaxies with respect to the matter distribution. However as one probes to smaller scales, the relationship of the underlying DM field to the observed galaxies becomes highly non-trivial. Not only must the bias of DM halos themselves be accounted for, but poorly understood galaxy formation and feedback processes also become prominent. The unknown processes governing the relationship between galaxies and their host halos is known as the galaxy-halo connection. This ambiguity causes a direct effect on halo-scale clustering but will also propagate to the larger, cosmologically-relevant scales. In order to counteract this effect, the power spectrum models are equipped with non-cosmological bias and nuisance terms intended to absorb any uncertainty in the knowledge of processes at small scales. With the Effective Field Theory of Large Scale Structure (EFTofLSS; \citep{Baumann2012, Carrasco2012, Ivanov2022}), the physics on scales smaller than a given cutoff are coarse-grained into a few ``effective'' parameters. However, quantifying the performance of these parameters to absorb changes in the galaxy-halo connection is essential.

% Motivation
The sheer volume of data that DESI collects presents new challenges for theoretical modelling and the control of systematics. The effect of the galaxy-halo connection on the compressed 2-point parameters for the extended Baryon Oscillation Spectroscopic Survey (eBOSS) using template-based methods was investigated in \citep{Rossi2021}. With increased volume and the addition of the ELG tracer, this sensitivity is greater for DESI. ELGs are young, star-forming galaxies, often occupying the satellite regions of halos. Hence, their clustering is more dominated by the complex processes of galaxy formation than other tracers. The effect of variations in the galaxy-halo connection for ELGs on the cosmological parameters recovered with EFT models has not yet been fully explored. The EFT models used in the DESI Full-Shape analysis have been rigorously tested in \cite{KP5s1-Maus, KP5s2-Maus, KP5s3-Noriega, KP5s4-Lai}. These papers validate the performance of the models into the mildly non-linear regime ($k\sim 0.2\hMpc$) by comparing them to mock data created assuming a simple, fixed galaxy-halo connection model. This work utilises the Halo Occupation Distribution (HOD; \cite{Berlind2002}) framework to explore the modelling robustness to a wider variety of galaxy-halo connection models. The ``HOD-dependent systematic error" is defined as any additional contribution to the uncertainty as a result of varying the galaxy-halo connection. This can be quantified at the level of the cosmological parameters as was done for the DESI 2024 BAO analysis, detailed in \cite{KP4s10-Mena-Fernandez} and \cite{KP4s11-Garcia-Quintero}, or following a new method at the level of the data vector proposed in this work. We explore how two HOD-dependent effects---(i) the ability of the EFT model to marginalise over small-scale effects and recover unbiased cosmological parameters and (ii) the additional contribution of the DESI 2024 Full-Shape nuisance term priors relative to the likelihood, which we refer to as the `prior weight effect'---contribute at the level of the data vector and compare this to a parameter-level-based estimate. To include the systematic contribution at the level of the data vector, we build a covariance matrix from mocks following two different approaches:
\begin{itemize}
    \item Isolating the cosmologically-relevant uncertainty in the power spectrum that arises from the inability of the EFT model to capture small-scale physics, closely mirroring the parameter-level method.
    \item Directly quantifying the HOD-dependent variation of mock data vectors.
\end{itemize}
Both of these approaches describe an extra effective contribution to the data covariance matrix, in contrast to the more intuitive approach of inflating uncertainties on cosmological parameters. This ensures the estimated HOD-dependent systematic uncertainty is independent of combinations with external datasets (e.g. BAO, cosmic microwave background probes, etc.). While we demonstrate that these methods propagate equivalent uncertainty to the parameter posteriors in a \LCDM\ scenario, the method of directly quantifying the variation of the data vectors should be more general in terms of the choice of model and freedom of parametrisation.

% What I am going to show
The paper is organised as follows. \cref{sec:galaxy-halo} motivates the necessity for exploring a wide range of models for the galaxy-halo connection and details the HOD models used in this analysis. 
In \cref{sec:data}, we describe the suite of mocks and covariance matrices used to explore the HOD-dependency of the Full-Shape fit.
In \cref{sec:method}, we discuss the power spectrum model, fitting method and describe our approach for including HOD-dependent systematics at the level of the data vector.
\cref{sec:results} discusses the validation of our method and the impact of our results for the DESI DR1 analysis.
In \cref{sec:conclusions}, we summarise our findings and highlight their implications for future analyses.

\section{The galaxy-halo connection}
\label{sec:galaxy-halo}
Large cosmological N-body simulations allow the distribution of DM to be studied in great detail (see \citep{Kuhlen2012} for a review). However, understanding the distribution of galaxies is key in order to compare to observable quantities. Often the DM distribution alone will be simulated due to the large computational cost of a full hydrodynamical simulation and hence some additional prescription to map from DM halos to galaxies is required. One such framework is the HOD---a probabilistic model that aims to encapsulate the complex physics of galaxy formation \citep{Vogelsberger2020} in a small number of empirically tuned parameters. In its simplest form, the probability that a halo with properties $\mathbf{X}$ will host $n$ galaxies, $P(n|\mathbf{X})$, is predominately driven by the mass of the halo, $M_h$ \cite{Zheng2005}. However, other non-local factors---known as assembly bias---can be included to better match observations \cite{Alam2020}. Exploring the HOD parameter space allows two distinct effects to be probed:
\begin{itemize}
    \item Uncertainty in the knowledge of the galaxy-halo connection imparted by the variety of different HOD forms and parameter values.
    \item Uncertainty in the randomness of galaxy formation imparted by the stochastic nature of sampling the distribution.
\end{itemize}
A variety of HOD models describing both the central and satellite galaxy occupations for each DESI tracer are used in this work. The models, summarised in \cref{tab:HODs}, were explored in previous work. We direct the reader to the references provided for additional details. Using the \Abacus simulations described in \cref{sec:data}, these models are tuned to approximately reproduce the clustering of the DESI One-Percent Survey \citep{DESI2023b.KP1.EDR} on small scales. This high-completeness sub-sample of the full DESI volume provides extremely accurate measurements of the small-scale clustering, ideal for investigating the galaxy-halo connection. In general, each HOD model is fit to the small-scale clustering with cosmology fixed to that of the base \LCDM\ \Abacus simulations, and the posterior distribution is used to determine the best-fit HOD parameter values. However, the specifics of the method for each tracer are detailed below.

\subsection{\label{sec:HOD_LRG}HOD models for LRG}
HOD models for DESI LRGs were implemented using the \texttt{AbacusHOD} code \cite{Yuan2022} and are detailed in \citep{Yuan2024}. Following the work of \citep{KP4s10-Mena-Fernandez}, we explore a selection of 8 models: 4 variations of the baseline model (denoted as the `A' models) and 4 extended `B' models. The best-fit models in each class are numbered 0 while 3 additional variations, numbered 1 to 3, randomly sample the posterior around the best-fit HOD parameters in each case.

In the A models, galaxies populate halos of mass $M_h$ according to \cite{Zheng2007} where the mean occupation numbers of a given halo for centrals and satellites are given by
\begin{align}
    \bar{n}_{\mathrm{cent}}^{\mathrm{LRG}}(M_h) & = \frac{f_\mathrm{ic}}{2}\,\mathrm{erfc} \left[\frac{\log_{10}(M_{\mathrm{cut}}/M_h)}{\sqrt{2}\sigma}\right], \label{eq:zheng_hod_cent}\\
    \bar{n}_{\mathrm{sat}}^{\mathrm{LRG}}(M_h) & = \left[\frac{M_h-\kappa M_{\mathrm{cut}}}{M_1}\right]^{\alpha}\bar{n}_{\mathrm{cent}}^{\mathrm{LRG}}(M_h).
    \label{eq:zheng_hod_sat}
\end{align}
Mass thresholds $M_{\mathrm{cut}}$ and $\kappa M_\mathrm{cut}$ set the minimum halo mass to host a central galaxy and satellite galaxy, respectively. $M_1$ is approximately the typical mass of a single-satellite-hosting halo. The transition from empty to central-hosting halos is dictated by the value of $\sigma$ while the exponent $\alpha$ controls the slope of the satellite occupation distribution. A downsampling factor $f_\mathrm{ic}$, where $0 < f_\mathrm{ic}\leq 1$, is included to account for survey incompleteness. With a further 2 parameters that bias the galaxy velocities relative to that of the host halo, a total of 8 parameters can be tuned to match the observed clustering. 

The B model additionally accounts for assembly bias with 2 environment-dependent parameters, $B_\mathrm{cent}$ and $B_\mathrm{sat}$,  by modulating galaxy formation based on the local density. A final parameter, $s$, that modifies the radial distribution of satellites within the halo is included in order to capture some baryonic effects. These additional parameters are defined in Eqs. 10, 11 and 7 of \cite{Yuan2022}, respectively. As a result, this model has 11 parameters.

Central galaxies were assigned the halo centre position and velocity by sampling a Bernoulli distribution with a mean equal to $\bar{n}_{\mathrm{cent}}^{\mathrm{LRG}}$. The assignment of satellites was similar but they instead follow a Poisson distribution with the positions and velocities assigned randomly to a host particle belonging to the halo. To create mocks for clustering analyses, the parameters were tuned to match the 2D correlation function, $\xi(r_p, \pi)$, measured in the One-Percent Survey, where $r_p$ and $\pi$ are the galaxy pair separation components perpendicular and parallel to the line-of-sight, respectively. The optimisation was performed in the range $0.1 \Mpch < r_p,\,\pi < 30 \Mpch$, and included an additional constraint on number density that allows for sample incompleteness, while penalising HOD models that produce insufficient number densities. Model A0 provides the best fit to the data.

\subsection{\label{sec:HOD_ELG}HOD models for ELG}
HOD models for DESI ELGs are detailed in \cite{Rocher2023} and \cite{Yuan2023}. We explore 21 different models following those used in \cite{KP4s11-Garcia-Quintero}. The baseline models for central galaxies are summarised below:
\begin{itemize}
    \item \textbf{GHOD}: Gaussian distribution around a logarithmic mass mean.
    \item \textbf{SFHOD}: Asymmetric star forming model with a decreasing power law for high mass halos.
    \item \textbf{HMQ}: High Mass Quenched model in which a quenching parameter controls the central occupation probability of high mass halos.
    \item \textbf{mHMQ}: Modified HMQ model with quenching parameter set to infinity.
    \item \textbf{LNHOD$_1$}: Log-normal model.
    \item \textbf{LNHOD$_2$}: Log-normal model tuned to smaller scale clustering.
\end{itemize}
The central galaxies sample a Bernoulli distribution and were assigned the position and velocity of the halo centre. The satellite galaxies sample a Poisson distribution and were positioned according to a Navarro-Frenk-White (NFW; \cite{Navarro1996}) profile. The satellite galaxy velocities are normally distributed around their mean halo velocity with a dispersion equal to that of the halo particles, rescaled by an extra free parameter that accounts for velocity biases.

The 6 baseline models can then be combined with a number of extensions. We explore 9 extended models that incorporate various permutations of the following effects:
\begin{itemize}
    \item Concentration-based assembly bias (\textbf{C}):  halo occupation is modulated by the halo concentration.
    \item Environment-based assembly bias (\textbf{Env}): halo occupation is modulated by the local density.
    \item Shear-based assembly bias (\textbf{Sh}): halo occupation is modulated by local density anisotropies.
    \item Modified satellite profile (\textbf{mNFW}): satellite galaxies follow a modified NFW profile that includes an exponential term.
    \item Galactic conformity (\textbf{cf}): satellite galaxies only occupy halos with a central galaxy.
    \item No 1-halo term contribution (\textbf{1h}): halos are only occupied by a single galaxy.
\end{itemize}
The models were implemented using a method based on Gaussian processes, with fixed number density of around $2.5\times 10^{-3}\hhhMpc$, as detailed in \cite{Rocher2023b}, and were tuned to jointly fit the projected correlation function, $w(r_p)$, and the correlation function monopole and quadrupole, $\xi_0(s)$ and $\xi_2(s)$, respectively, of the One-Percent Survey. The models were fit to $w(r_p)$ in the range $0.04 \Mpch < r_p < 32 \Mpch$ with $\pi_\mathrm{max} = 40 \Mpch$ used for the line-of-sight integration. The correlation function multipoles were fit up to $s=32\Mpch$, with smaller scales ($s_\mathrm{min}= 0.17\Mpch$) included in fits using the mHMQ and LNHOD$_2$ models than for other models ($s_\mathrm{min}= 0.8\Mpch$). Model mHMQ+cf+mNFW provides the best fit to the data.

Additionally, six high mass quenched models, denoted $\textbf{HMQ$^{(3\sigma)}_i$ ($i=1,2,...,6$)}$, created with \texttt{AbacusHOD} are explored \citep{Yuan2023}. As with the LRGs, positions and velocities were assigned to centrals using the halo centre and to satellites using random particles within the halo. The models sample the posterior around the best-fit HOD parameters and include velocity bias for both centrals and satellites. Models $i=4,5,6$ also include a complex prescription of galaxy conformity. The models were tuned to the 2D correlation function, $\xi(r_p,\pi)$, of an early version of the One-Percent Survey in the range $0.04 \Mpch < r_p < 32 \Mpch$ with $\pi_\mathrm{max} = 40 \Mpch$. As with the LRGs, an additional number density constraint was imposed on the fitting procedure.

\subsection{\label{sec:HOD_BGS}HOD models for BGS}
HOD models for the magnitude-limited DESI BGS are detailed in \cite{Smith2024}. The models very closely resemble those of the LRGs, but the occupation numbers are instead defined as smooth functions of luminosity, $L$, (i.e. $\bar{n}^{\mathrm{BGS}}(>L | M_h)$) in order to correctly reproduce the clustering for any given magnitude-limit. Additionally, the error function used to model the mass step in \cref{eq:zheng_hod_cent} is converted to a pseudo-Gaussian in order to prevent unphysical crossing of HOD samples with different absolute magnitude thresholds. These models were tuned to the measured projected correlation function, $w(r_p)$, of the One-Percent Survey integrated to $\pi_\mathrm{max} = 40 \Mpch$, with $0.1 \Mpch < r_p < 80 \Mpch$. 17 meta-parameters that control luminosity dependence of the HOD parameters were varied, with an additional constraint on the number density, in order to perform the optimisation. Central galaxies were populated following the Monte Carlo method outlined in \cite{Smith2017}, while satellites were sampled using a Poisson distribution and positioned according to an NFW profile. The satellite velocities were drawn from a normal distribution with a width that is related to properties of the host halo. 11 variations in HOD parameters, generated by sampling the posterior around the best-fit values, are explored. Model BGS$_0$ provides the best fit to the data, with the other variations denoted as BGS$_{1\text{-}10}$.

\subsection{\label{sec:HOD_QSO}HOD models for QSO}
HOD models for DESI QSOs, also detailed in \citep{Yuan2024}, are almost identical to the standard LRG models. Motivated by a lack of evidence for central-satellite correlation, the satellite distribution in \cref{eq:zheng_hod_sat} is modified by removing the dependence on the central galaxy through $\bar{n}_{\mathrm{cent}}^{\mathrm{LRG}}$. As with the LRGs, the clustering and number density of these models were tuned to the 2D correlation function, $\xi(r_p, \pi)$, of the One-Percent Survey, with 3 variations, QSO$_{1\text{-}3}$, that sample the posterior around the best-fit HOD parameters of model QSO$_0$ explored.

\begin{table}[]
    \centering
    \begin{tabular}{|c|c|c|c|p{6.4cm}|c|}
        \hline
        Tracer & $z$ & \# models & Fit to & Specification & Ref.\\ \hline
        BGS  & 0.2 & 11 & $w(r_p)$                         & Luminosity-varying HOD: Step function (centrals) + power law (satellites) & \cite{Smith2024}\\
        LRG  & 0.8 & 8  & $\xi(r_p, \pi)$                  & Step function (centrals) + power law (satellites) + assembly bias & \cite{Yuan2024}\\
        ELG  & 0.8 & 6  & $\xi(r_p, \pi)$                  & Quenching at high halo mass + galactic conformity & \cite{Yuan2023}\\
        ELG  & 1.1 & 15 & $w(r_p)$, $\xi_0(s)$, $\xi_2(s)$ & Quenching at high halo mass + assembly bias + additional modifications & \cite{Rocher2023}\\
        QSO  & 1.4 & 4  & $\xi(r_p, \pi)$                  & Step function (centrals) + central-independent power law (satellites) & \cite{Yuan2024}\\
        \hline
    \end{tabular}
    \caption{Summary of the HOD ensemble explored. The tracer, redshift, number of models, One-Percent survey summary statistic fit to, model specifications and corresponding reference are listed. $w(r_p)$ denotes the projected correlation function, $\xi(r_p, \pi)$ denotes the 2D correlation function and the correlation function monopole and quadrupole are denoted by $\xi_0(s)$ and $\xi_2(s)$, respectively.}
    \label{tab:HODs}
\end{table}

\section{Data}
\label{sec:data}
\subsection{\Abacus HOD mock catalogues}
\label{sec:mocks}

We employ the suite of 25 base-\LCDM\ \Abacus N-body simulations \citep{Maksimova2021, Garrison2019, Garrison2021} to test HOD-dependent effects. These high-precision simulations are constructed by evolving $6912^3$ DM particles in a cubic box of volume $(2\Gpch)^3$. In what follows, we refer to this volume as `V1'. The simulations are generated using a cosmology according to the mean estimates of the \emph{Planck} 2018 TT,TE,EE+lowE+lensing posterior: $\omega_\mathrm{cdm} = 0.1200$, $\omega_\mathrm{b} = 0.02237$, $\sigma_8 = 0.811355$, $n_s = 0.9649$, $h = 0.6736$, $w_0 = -1$, $w_a = 0$ and a single $0.06$ eV massive neutrino \citep{Planck2018}. Halos are identified with the \textsc{compaso} algorithm \citep{Hadzhiyska2022} at a redshift snapshot of interest and populated with the HOD models outlined in \cref{sec:galaxy-halo}. Snapshots are selected at $z=$ 0.2, 0.8 and 1.4 for the BGS, LRG and QSO samples, respectively. For the ELG sample, two different snapshots are explored. The six HMQ$^{(3\sigma)}_i$ models based on \texttt{AbacusHOD} are used to populate a snapshot at $z=0.8$ and all other models at $z=1.1$. This leads to 200 mocks for LRG, 525 mocks for ELG, 100 mocks for QSO and 11 mocks for BGS (only mocks derived from a single \Abacus realisation are available for BGS).

\subsection{Power spectrum measurements}
\label{sec:pks}
Power spectrum measurements for each of the HOD cubic mocks are provided in \cite{KP4s10-Mena-Fernandez} and \cite{KP4s11-Garcia-Quintero}. The measurements are computed using the DESI package \texttt{pypower}\footnote{\url{https://github.com/cosmodesi/pypower}} adopting the periodic box estimator \cite{Hand2017} in which multipoles are calculated according to 
\begin{equation}
    P_\ell(k)=\frac{2\ell+1}{V}\int\frac{d\Omega_k}{4\pi}\delta_g(\vec{k})\delta_g(-\vec{k})\mathcal{L}_\ell(\mu)-P_\ell^{\rm shot\text{-}noise}.
    \label{eq:ps_multipoles}
\end{equation}
Here, the galaxy overdensity is denoted by $\delta_g\equiv n_g/\bar{n}_g - 1$, $V$ is the volume of the box, $\Omega_k$ is the solid angle in Fourier space, $\mathcal{L}_\ell$ are the Legendre polynomials of order $\ell$ and $\mu$ is the cosine of the angle between wavevector $\vec{k}$ and the line of sight. The Poisson shot-noise term is only subtracted for the monopole ($\ell=0$). To estimate the power spectrum, the density field is interpolated on a $512^3$ mesh created using a triangular-shaped cloud prescription. The measurements are computed from $k=0-0.2\hMpc$ with a binning of $\Delta k=0.001\hMpc$. For comparison to theory, the measurement is then re-binned with a spacing of $\Delta k=0.005\hMpc$.

\subsection{DR1-like data vectors}
\label{sec:cut-sky}

\begin{figure}
    \centering
    \includegraphics[width=0.8\linewidth]{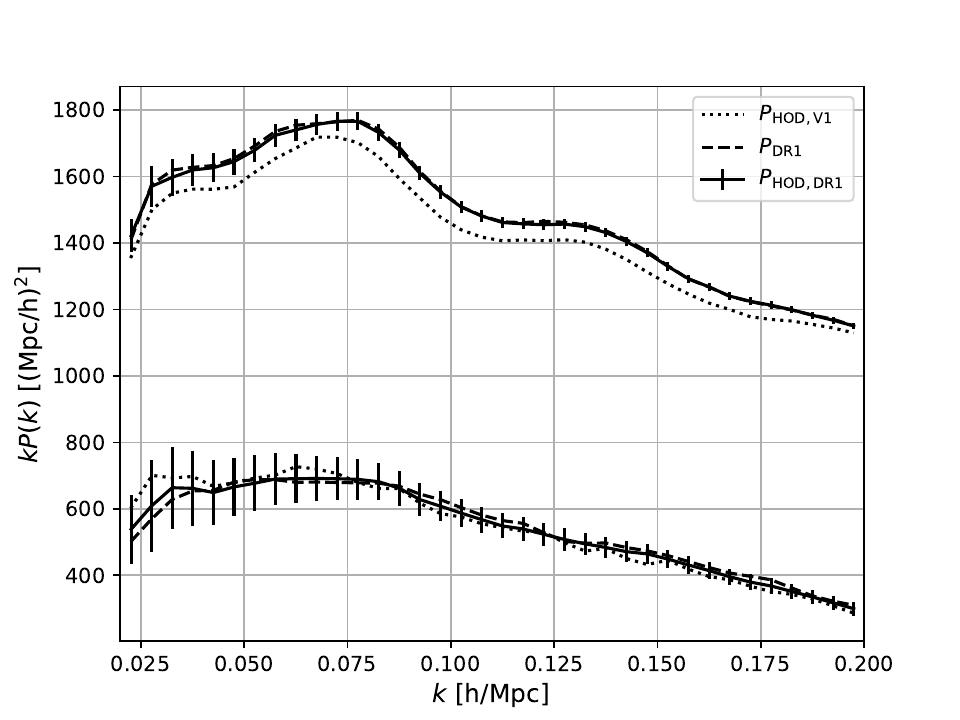}
    \caption{Validation of the DR1-like power spectrum (\emph{solid}) for the best-fit LRG HOD model, A0. The DR1-like power spectrum is created by applying the DR1 window to the cubic mock (\emph{dotted}). The DR1 window has been created with FFA effects included and $\theta$-cut applied. The true DR1 mock (\emph{dashed}) is provided for comparison. Uncertainties are determined using a DR1 Gaussian covariance.}
    \label{fig:DR1-like-pk}
\end{figure}

Throughout the rest of the paper, we will use the following terminology when denoting the type of data vector used:
\begin{itemize}
    \item \textbf{Cubic}: Power spectrum measured on individual realisations of the \Abacus cubic box populated with all HOD models. Fits to this power spectrum are performed with the V1 covariance---the analytic covariance corresponding to a single $(2 \Gpch)^3$ cubic box.
    \item \textbf{DR1-like}: Mean of 25 power spectrum measurements from individual realisations of the \Abacus cubic box convolved with the DR1 window. These are generated for each HOD model. Fits to this power spectrum are performed with the analytic covariance corresponding to the DR1 volume.
    \item \textbf{Fixed HOD DR1}: Power spectrum measured on mocks that incorporate survey geometry and selection effects. Only available for a \emph{single} HOD model corresponding to the one that represents the best match to the DESI DR1 clustering.
\end{itemize}
In this section, we describe how the `DR1-like' data vectors are generated. More details on the cubic and fixed HOD DR1 mocks can be found in Section 11 of \cite{DESI2024.II.KP3}. The analytic covariance matrices are described in the following section.

In order to explore relevant HOD-dependent effects for DR1, realistic mock data is a requirement (more discussion on this is in \cref{sec:systematic}). Mocks that incorporate DR1 survey geometry and selection effects have been generated with a single, fixed HOD and utilised extensively for systematic tests \cite{KP3s8-Zhao}. While a variety of possible systematic sources have been explored using these realistic survey mocks, they do not permit tests related to varying the galaxy-halo connection. To explore HOD-dependent systematics in the context of DR1, rather than taking the computationally expensive approach of creating mocks with survey effects included for each HOD model, we instead produced what we will refer to as `DR1-like' data vectors. These DR1-like measurements are generated directly from the power spectrum measured on cubic mocks and only require a simple window convolution to mimic the full mock-based approach. The mean of 25 power spectrum measurements obtained from individual realisations of the cubic mocks was used to create a ``mock theory'' vector, $\hat{P}_{\ell^{\prime}}^t(k^{\prime})$, which was then convolved with the realistic DR1 window matrix, \tens{W}, that captures the effect of fibre assignment estimated using the `fast-fibreassign' method (FFA; \cite{KP3s6-Bianchi, Hanif2024}) and the effect of the $\theta$-cut. The window matrix is estimated using a catalogue of random positions following the method detailed in Section 10.1.2 of \cite{DESI2024.II.KP3}. The random catalogue spans the survey footprint of the chosen tracer and samples the selection function of the data to ensure that no spurious clustering signal is measured when estimating the power spectrum. The random sample is subject to the FFA algorithm which is used to efficiently emulate the statistical effect of the probabilistic assignment of DESI fibres to the target galaxy sample. The $\theta$-cut, discussed in \cite{KP3s5-Pinon}, is imposed to mitigate fibre assignment effects by removing pairs at small angular separations. It thus induces a sensitivity of the window to high-$k$ modes and therefore a basis rotation has also been applied in order to increase the compactness of the window. In Fourier-space, this convolution takes the form
\begin{equation}
    \hat{P}_\ell^\mathrm{DR1}(k) = W_{\ell \ell^{\prime}}(k, k^{\prime}) \hat{P}_{\ell^{\prime}}^t(k^{\prime}),
    \label{eq:window}
\end{equation}
where $k^\prime$ and $k$ denote the input and output wavenumbers with maximum values $k^\prime=0.35 \hMpc$ and $k=0.2 \hMpc$, respectively. The input multipoles extend to the hexadecapole, $\ell^\prime=(0,2,4)$, while the output was computed only up to the quadrupole, $\ell=(0,2)$. These measurements are not fully realistic realisations of the mock measurements (they rely on the accuracy of the window matrix) but are inexpensive to produce and could therefore be generated for each LRG, ELG and QSO HOD model. DR1-like data vectors were not generated for the BGS tracer as we did not have the required number of cubic mock realisations to reduce sample variance in estimating $\hat{P}_{\ell^{\prime}}^t(k^{\prime})$. The window matrices correspond to a redshift binning with limits $z=[0.6,0.8], [1.1,1.6], [0.8,2.1]$ for the LRG, ELG and QSO samples, respectively. While the DESI DR1 analysis uses three LRG redshift bins, only a single DR1-like data vector was generated due to the computational cost required to produce covariance matrices for each redshift bin and HOD model (see \cref{sec:cov_mat}). A single bin per tracer is sufficient for the purpose of this work, given that we do not expect the effect of non-cosmological (nuisance) parameter priors, investigated using these data vectors, to change significantly across bins of a given tracer. The fitting process, described in \cref{sec:model}, scales stochastic term priors by the shot-noise of the input data. To ensure that this scaling is correct relative to DR1 data (not the cubic box input $P(k)$), the newly-generated, DR1-like \texttt{pypower} power spectrum files were assigned a shot-noise value equal to that of the fixed HOD DR1 mocks. \cref{fig:DR1-like-pk} demonstrates excellent agreement between the DR1-like data vector for the best-fit LRG model, A0, and the fixed HOD DR1 mock.

\subsection{Covariance matrices}
\label{sec:cov_mat}

For the DR1 analysis, covariance matrices are constructed from 1000 effective Zel'dovich approximate mock simulations (\texttt{EZmocks}; \citep{EZmock2015}). These large $(6\Gpch)^3$ mocks allow the survey geometry of the DR1 sample to be reproduced without replication of the simulation box. The \texttt{EZmocks} are generated, using an effective biasing scheme, to match the clustering of the \Abacus simulations with a fixed galaxy-halo connection model. Therefore, \texttt{EZmock}-based covariance matrices do not account for the difference in clustering amplitude that arises when varying the HOD (see \cref{fig:HODs}). In this work, we aim to produce covariance matrices that are tuned to the mock clustering for each HOD model and, in the case of fits to the DR1-like data, ensure that the covariance amplitude is scaled to that of the data. For this reason, we compute analytic Fourier-space covariance matrices for each HOD model using the DESI analytical covariance code \texttt{thecov} \citep{KP4s8-Alves}.\footnote{\url{https://github.com/cosmodesi/thecov}} This code follows the groundwork of \citep{Wadekar2019} and \citep{Kobayashi2023} allowing the computation of power spectrum covariance matrices in arbitrary geometries. The covariance of the power spectrum fundamentally depends on the 4-point correlator. Using Wick's theorem, this can be decomposed into products of 2-point correlators---the Gaussian contribution---and a trispectrum term that is non-zero in the presence of non-Gaussianity. We neglect the non-Gaussian terms for simplicity in the case of the V1 covariance for the cubic mocks, given that these have a marginal contribution to the cosmological posteriors \cite{Wadekar2020}.

Covariance matrices are generated for both the V1 and DR1 volumes. The performance of the analytic covariance is validated against the \texttt{EZmock} approach in \cite{KP4s6-Forero-Sanchez}, who find that the variance is slightly lower than that observed in the \texttt{EZmocks}. However, the performance of the analytic covariance is more than sufficient for maximisation of the likelihood or posterior as conducted in this work. Section 10.2 in \cite{DESI2024.II.KP3} details that the \texttt{EZmocks} are unable to reproduce the variance of the real data, due to shortcomings in the FFA approximation. In order to account for this, a scale-independent rescaling factor is applied to the \texttt{EZmock}-derived covariance matrix for each tracer based on their mismatch with the configuration-space DR1 covariance \cite{KP4s7-Rashkovetskyi}. As the amplitude of the analytic covariance is similar to that of the \texttt{EZmocks}, these rescaling factors must also be applied here, in order to match the variance of the data. The factors, listed in Table 8 of \cite{DESI2024.II.KP3}, are $1.39$, $1.15$, $1.29$ and $1.11$ for the BGS, LRG, ELG and QSO redshift bins explored in this work, respectively.

As the cubic box mocks are not subject to survey geometry or selection effects, the V1 covariance is easily generated by passing the estimated power spectrum as input to \texttt{thecov}. In contrast, a catalogue of random positions must also be provided when estimating the covariance for the DR1-like HOD mocks in order to correctly account for the survey window. The random catalogues are the same as those used for estimating the window matrices in \cref{sec:cut-sky}. Once the analytic covariance matrix, $\tens{C}$, has been produced, the rotation to increase the compactness of the window function (again following Section 10.1.2 of \cite{DESI2024.II.KP3}) is applied,
\begin{equation}
    \tens{C}^\prime = \tens{M} \tens{C} \tens{M}^T,
\end{equation}
where rotation matrix \tens{M} is determined by an optimisation
procedure according to \cite{KP3s5-Pinon}. The rotation is applied for consistency with the DR1 analysis pipeline but has minimal effect on the results. To reproduce the variance of the data, we correct the matrices with the corresponding rescaling factors, as discussed earlier. Additionally, a factor of $1.5$, obtained by roughly matching to the DR1 \texttt{EZmock} covariance, is applied to the ELG covariance in order to account for a discrepancy due to the fact that the redshift of the input cubic box power spectrum ($z=1.1$) does not match the redshift of the \texttt{EZmocks} ($z=1.325$) and hence that of the data. These factors are applied consistently to all HOD models of a given tracer.

\section{Method}
\label{sec:method}

The Full-Shape analysis performed in BOSS and eBOSS (e.g., \cite{Alam-BOSS:2017,Gil-Marin2020, Bautista2021}) was primarily based on a template-fitting approach (although other methods have been used, e.g. \cite{Sanchez2017}) which provided constraints on a set of summary parameters, a form of data compression. Cosmological results were then obtained in a subsequent step, through fitting models to the summary parameters assuming a Gaussian likelihood.\footnote{A similar approach is also naturally applied in BAO fits, where results are expressed in terms of BAO scaling parameters, $\alpha_\perp$ and $\alpha_\parallel$, as done in \cite{DESI2024.III.KP4}, and then interpreted in a cosmological context, as in \cite{DESI2024.VI.KP7A}.} This two-step process lent itself to expressing systematic error contributions in the form of an additional uncertainty in the results for the compressed parameters, which can be added in quadrature to the statistical errors and thus automatically propagated to cosmological parameter results in any model or in combination with any external data.

However, as detailed in \cite{DESI2024.VII.KP7B}, for the DESI DR1 results we use a full EFT-based approach, referred to as Full Modelling, in which cosmological parameters are fit directly from the data, in preference to the two-step template-based compression. While this has many benefits, it complicates the inclusion of possible systematic error contributions to the final error budget at the level of the parameters as before, since recovered parameter values depend both on the choice of which parameters are varied in the analysis and which external datasets, if any, are included in the fit alongside the galaxy power spectrum. Adding systematic error contributions at the parameter-level as before would thus require a separate estimate of the systematic uncertainty for each cosmological model and each combination of datasets that is to be considered---a prohibitive task.

Therefore, we propose to take a different approach: we quantify the effects of the systematic errors in terms of an additional \emph{effective} uncertainty at the level of the power spectrum data vector, as explained in \cref{sec:systematic} below. This is expressed in the form of an additional covariance matrix contribution, $\tens{C}_{\rm sys}$, where the subscript here reflects that the source of this contribution is systematic. $\tens{C}_{\rm sys}$ is to be added (together with any other similar contributions from other sources) to the covariance matrix $\tens{C}_{\rm stat}$ representing the statistical measurement uncertainties in the power spectrum when performing a fit to the data. This additional contribution to the covariance is not a true reflection of the level of uncertainty in capturing variations in the galaxy-halo connection, but rather an effective contribution that correctly propagates the true uncertainty to the marginalised parameter posteriors. Such an approach is then generally applicable irrespective of which model parameters are held fixed or varied, or which additional datasets are included in the fits.

\subsection{Model}
\label{sec:model}

All modelling and fitting routines used in this work are included within the DESI pipeline for likelihood analysis, \texttt{desilike}.\footnote{\url{https://github.com/cosmodesi/desilike}} We use the implementation of the \Velocileptors Lagrangian Perturbation Theory (LPT) code \cite{Chen2020, Chen2021} as our choice of EFT model to compute the redshift-space power spectrum monopole and quadrupole following the baseline parametrisation of \cite{DESI2024.V.KP5}. This choice is arbitrary given the consistency of the DESI EFT codes \cite{KP5s1-Maus}. The EFT models employ perturbation theory, in combination with a course-graining of small-scale physics, in order to provide a rigorous prescription of the galaxy power spectrum into the mildly non-linear regime. The perturbative part, $P_\mathrm{s,g}^\mathrm{PT}$, is computed to next-to-leading order, also known as the 1-loop contribution. Complicated small-scale physics are ``integrated out'' of the perturbative theory into a few counterterms and stochastic parameters which are added at leading order (tree-level). The counterterms, $\alpha_n$, capture the coupling of small-scale physical processes to the larger scales of interest while the stochastic terms, $\mathrm{SN}_n$, account for random, uncorrelated small-scale fluctuations. As galaxies are biased tracers of the underlying DM distribution, a Taylor expansion of the galaxy overdensity in terms of the matter overdensity propagates Lagrangian bias terms (related but not equivalent to the Eulerian ones, e.g. $b = 1 + b_1$) into the final result. This leads to the redshift-space power spectrum, 
\begin{equation}
    P_\mathrm{s,g}(k,\mu) = P_\mathrm{s,g}^\mathrm{PT}(k,\mu, b_1, b_2, b_s) + (b+f\mu^2)(b\alpha_0 + f\alpha_2\mu^2)k^2 P_\mathrm{s,lin}(k,\mu) + (\mathrm{SN}_0 + \mathrm{SN}_2 k^2\mu^2),
\end{equation}
where $P_\mathrm{s,lin}$ is related to the linear power spectrum, $f$ is the linear growth rate and $\mu$ is the cosine of the angle between wavevector $k$ and the line of sight. We urge the reader to refer to \cite{KP5s1-Maus,KP5s2-Maus} for further details of the model formalism and validation against \Abacus mocks with a fixed HOD model.

Allowing $b_1$ (linear), $b_2$ (quadratic) and $b_s$ (shear) bias terms to vary grants maximum flexibility of the model to marginalise over uncertainties in the galaxy-halo connection. The third order bias term $b_3$ is fixed due to degeneracies with the counterterms following \cite{DESI2024.V.KP5}. The values of the additional counterterm and stochastic parameters are not known a priori but they can be constrained by the data in addition to the cosmological parameters. Furthermore, their rough magnitude can be estimated from theory or simulations allowing reasonable `physically motivated' priors, discussed in detail in \cite{KP5s2-Maus}, to be placed on them. In this basis, counterterms scale relative to the linear theory multipoles and stochastic terms scale with the Poissonian shot-noise and the characteristic halo velocity dispersion. The baseline for the DESI Full-Shape analysis investigates five cosmological parameters---although little information can be gained from the baryon density, $\omega_b$, as it is not constrained by the data and requires a prior which, in this work, is derived from Big Bang Nucleosynthesis (BBN) constraints \cite{Schoneberg2024}. For this reason, we choose to exclude $\omega_b$ from any figures. The loose prior on the scalar spectral index, $n_s$, is a Gaussian centred at $n_s=0.9649$ with a width chosen to be $10\times$ the posterior uncertainty from \emph{Planck} \cite{Planck2018}. Prior choices for all 12 varied parameters are listed in \cref{tab:priors}. In addition to the default Gaussian priors on nuisance terms, flat priors were explored in order to investigate potential biases due to the imposition of our `physically motivated' priors. We refer to this prior cases as `uninformative' in order to differentiate it from the baseline `physical' case.

The \Velocileptors model was emulated within the \texttt{desilike} framework using a fourth-order Taylor expansion to increase the computational efficiency of the fitting procedure. As described in the next section, maximum a posteriori (MAP) estimates were obtained for the HOD ensemble by fitting to the monopole and quadrupole measurements of the \Abacus mocks over the range $k=0.02-0.2\hMpc$. These MAP estimates were determined using the \texttt{desilike} wrapper of the \texttt{Minuit} profiler \cite{Minuit}. We choose not to employ Markov Chain Monte Carlo (MCMC) sampling when computing the HOD systematic contribution to avoid the inclusion of projection effects (see Section 4.5 of \cite{DESI2024.V.KP5} for a discussion on this effect) which affect the posterior mean but not the MAP value. However, in figures where we compare MAP values to the marginalised posterior, we employ the Hamiltonian Monte Carlo sampling algorithm \texttt{NUTS} \cite{NUTS1,NUTS2} to compute this. In this case, the linear nuisance parameters of our model, $\alpha_n$ and $\mathrm{SN}_n$, have been analytically marginalised to accelerate sampling.

\begin{table}[]
    \centering
    \begin{tabular}{|c|c| |c|c|}
        \hline
        Cosmological & Prior & Nuisance & Prior\\ \hline
        $\omega_\mathrm{cdm}$   & $\mathcal{U}$[0.01, 0.99]         & $(1+b_1)\sigma_8$       & $\mathcal{U}$[0, 3] \\
        $\omega_\mathrm{b}$     & $\mathcal{N}$[0.02237, 0.00055]   & $b_2\sigma_8^2$       & $\mathcal{N}$[0, 5] \\
        $h$                     & $\mathcal{U}$[0.1, 10]            & $b_s\sigma_8^2$       & $\mathcal{N}$[0, 5] \\
        ln($10^{10}A_s$)        & $\mathcal{U}$[1.61, 3.91]         & $\alpha_0$  & $\mathcal{N}$[0, 12.5] \\
        $n_s$                   & $\mathcal{N}$[0.9649, 0.04]       & $\alpha_2$  & $\mathcal{N}$[0, 12.5] \\
                                &                                   & SN$_0$      & $\mathcal{N}[0,2] \times 1/\bar{n}_g$ \\
                                &                                   & SN$_2$      & $\mathcal{N}[0,5] \times f_{\rm sat} \sigma_{v}^2/\bar{n}_g$ \\
        \hline
    \end{tabular}
    \caption{\Velocileptors LPT varied parameters and priors used for fitting. The entries $\mathcal{U}$[min, max] and $\mathcal{N}$[$\mu$, $\sigma$] refer to uniform and Gaussian normal distributions, respectively. Non-cosmological (nuisance) priors have been applied according to a `physically motivated' parametrisation following \cite{KP5s2-Maus}. In this basis, counterterms scale relative to the linear theory multipoles and stochastic terms scale with the Poissonian shot-noise, $1/\bar{n}_g$, and the characteristic halo velocity dispersion, $f_{\rm sat} \sigma_{v}^2/\bar{n}_g$, where $f_{\rm sat}$ and $\sigma_{v}$ are the expected fraction and mean velocity dispersion of satellite galaxies, respectively. In the case of `uninformative' priors, infinite flat priors are instead imposed on nuisance parameters.}
    \label{tab:priors}
\end{table}

\subsection{Estimating the systematic contribution}
\label{sec:systematic}

This work aims to capture two independent contributions at the level of the power spectrum:
\begin{enumerate}
    \item The variation in clustering due to changing the HOD that cannot be marginalised over by varying the nuisance parameters of our model. Given that the mock-based statistical covariance is computed with a fixed galaxy-halo connection, this additional contribution covers uncertainty in allowing this connection to vary.
    \item The effect of nuisance parameter priors on the EFT model fit to a range of HOD mocks.
\end{enumerate}
In order to address the first point, we can generate a covariance matrix directly from the variation in the measured summary statistic of interest. For the purpose of this work, we focus on the power spectrum. Using the power spectrum measurements discussed in \cref{sec:pks}, we can calculate the residual, 
\begin{equation}
    \vec{\Delta P}^{\mathrm{A,B}}_i \equiv \vec{\hat{P}}^\mathrm{A}_{i} - \vec{\hat{P}}^\mathrm{B}_{i},
    \label{eq:residuals_true}
\end{equation}
of HOD models A and B for a given tracer at fixed mock realisation $i$. The power spectrum monopole and quadrupole in the range $k=0.02-0.2\hMpc$ with spacing $\Delta k=0.005\hMpc$ are concatenated such that $\vec{\Delta P}^{\mathrm{A,B}}_i$ is a vector of size $N_k=72$. By computing this residual, sample variance in the halo catalogue is eliminated by construction, while maintaining the HOD-dependent variance. This approach helps to isolate the contribution of the uncertainty in the galaxy-halo connection, which can otherwise be washed out when averaging each model over many mock realisations before comparing differences. \cref{eq:residuals_true} intrinsically contains some shot-noise due to the use of noisy individual mock realisations, however we verify that this contribution is negligible in \cref{sec:shotnoise}. While the magnitude of these residuals depends on the range of HOD models explored, \cref{fig:HODs} in \cref{sec:full_cov} demonstrates that we explore an extremely wide HOD prior space because only the small-scale clustering is matched. In this work, these measurements have been obtained with a fixed \LCDM\ cosmology but, in theory, variations across a wide range of cosmologies could be accounted for in an equivalent manner. However, we expect the cosmological dependency to be weak as \cref{eq:residuals_true} computes \emph{relative} shifts between power spectra of the HOD ensemble at a given cosmology. Additionally, altering the cosmology cannot drastically affect the HOD mock measurements because they must still roughly reproduce the observed clustering of the data.
From this, we compute the covariance matrix of the power spectrum residuals
\begin{equation}
    \tens{C}^{\rm V1}_\mathrm{HOD} = \frac{1}{N_\mathrm{perm} - 1}\Big(\vec{\Delta P} - \overline{\vec{\Delta P}}\Big) \Big(\vec{\Delta P} - \overline{\vec{\Delta P}}\Big)^T,
    \label{eq:C_HOD}
\end{equation}
where the set of residuals across all pairs of HOD models and mock realisations is given by
\begin{equation}
    \vec{\Delta P} \equiv \big\{ \vec{\Delta P}^{\mathrm{A,B}}_i \big\}_{\mathrm{A}\neq\mathrm{B}}\,\text{,   for $N_\mathrm{perm}$ permutations of A, B and $i$.}
    \label{eq:permutations}
\end{equation}

\begin{figure}
    \centering
    \includegraphics[width=\linewidth]{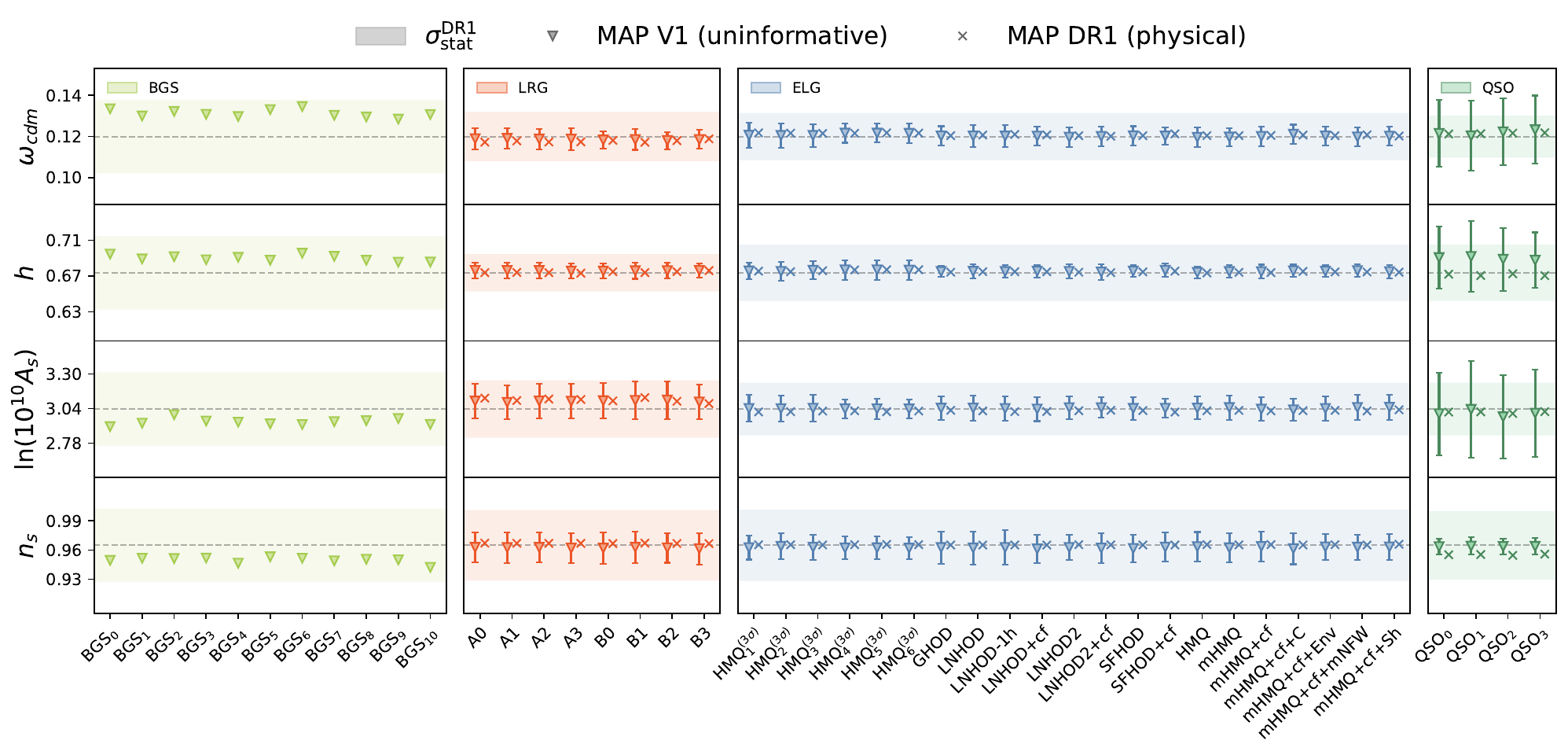}
    \caption{Best-fit cosmological parameters for different HOD models measured on cubic mocks with V1 covariance (V1) and DR1-like data with DR1 covariance (DR1). Fits to V1 have flat `uninformative' priors on nuisance terms such that differences between these points and those in the baseline `physical' parametrisation are due to the miscentering of priors---referred to as the `prior weight effect'. The coloured bands show the DR1 statistical uncertainty for each tracer at the redshift of the data (including \texttt{EZmock} rescaling). The error-bars show the standard deviation of 25 mock realisations. DR1-like data is generated from a power spectrum measurement averaged over 25 mocks, hence no error-bars are provided. Only one mock realisation is available for the BGS sample and so only one value corresponding to a single fit to V1 is shown.}
    \label{fig:shifts}
\end{figure}

This $N_k \times N_k$ covariance matrix captures all variation in the measured mock power spectra due to different galaxy-halo connection models within the wide, conservative HOD parameter space we explore, but by construction does not include any effects of sample variance in the underlying halo catalogues themselves (since differences are always computed at the same simulation realisation). The power spectrum of mocks populated with different HOD models may vary substantially in quantities not constrained by the small-scale clustering to which the models are fit such as the effective galaxy biases which produce the coherent amplitude shift seen in \cref{fig:HODs} of \cref{sec:full_cov}. This means that the term computed in \cref{eq:residuals_true} is, in general, large and leads to a covariance matrix with a highly non-diagonal structure. Although individual terms in this covariance can be significantly larger than those of the statistical covariance, $\tens{C}_{\rm stat}$, strongly correlated differences like this will mostly be accommodated within the bias and nuisance terms of the EFT model. Therefore, although $\tens{C}^{\rm V1}_\mathrm{HOD}$ defined as above would dominate the total covariance, $\tens{C}^{\rm V1}_\mathrm{tot} = \tens{C}^{\rm V1}_\mathrm{stat} + \tens{C}^{\rm V1}_\mathrm{HOD}$, the effect of including this term on the posterior constraints on cosmological parameters of interest will be small---indeed, if the model is flexible enough to perfectly accommodate such HOD variations without biases in the cosmological parameters, zero HOD contribution will be propagated to the cosmological posteriors. On the other hand, since the performance of the \Velocileptors model has only been validated on mock data generated with a single HOD model \cite{KP5s2-Maus,KP5s1-Maus} (and see also \cite{KP5s3-Noriega, KP5s4-Lai} for equivalent models), this extra covariance term allows us to incorporate any potential additional variations to cosmological parameter constraints that might arise in the context of other HOD scenarios. 

Finally, in order to investigate the effect on DR1 data, we apply the DR1 window matrix, $\tens{W}$, described in \cref{sec:cut-sky}, to the covariance matrix,
\begin{equation}
    C^{\rm DR1}_\mathrm{HOD}(k, k) = W(k, k^{\prime})  C^{\rm V1}_\mathrm{HOD}(k^\prime, k^\prime) W(k, k^{\prime})^\mathrm{T}.
    \label{eq:C_windowed}
\end{equation}
The convolution is computed up to $k^\prime=0.35\hMpc$ using the same input and output multipoles ($\ell$ and $\ell^\prime$) as in \cref{eq:window} although we have chosen not to denote them here for clarity. The covariance determined using \cref{eq:residuals_true} makes no reference to the specific theory model of the power spectrum, the choice of which cosmological parameters are varied in any fit or external datasets used to constrain the data. Hence, we refer to this method as the `general' approach given that it is valid for use in any of theory, parameter and dataset combination. However, $C^{\rm DR1}_\mathrm{HOD}$ is large and very far from diagonal in structure, which is inconvenient and leads to concerns about the numerical precision with which its elements can be determined from a small number of simulation realisations.

We therefore develop another alternative approach, in which the power spectrum residuals in \cref{eq:residuals_true} are instead replaced with
\begin{equation}
    \vec{\Delta P}^{\mathrm{A,B}}_i \equiv \vec{P}_{i}(\Vec{\Omega}_\mathrm{A},\Vec{n}_{\mathrm{bf}}) - \vec{P}_{i}(\Vec{\Omega}_\mathrm{B},\Vec{n}_{\mathrm{bf}}),
    \label{eq:residuals}
\end{equation}
where power spectra $\vec{P}_{i}(\Vec{\Omega}_\mathrm{X},\Vec{n}_{\mathrm{bf}})$ represent the theory prediction of the model evaluated at cosmological parameters, $\Vec{\Omega}_\mathrm{X}$, and nuisance parameters, $\Vec{n}_{\mathrm{bf}}$. $\Vec{\Omega}_\mathrm{X}$ is the MAP estimate of the cosmological parameters for the fit to the measured mock power spectrum for a given HOD model X. This best-fit estimate is determined for each HOD model A and B (with nuisance parameters free). $\Vec{n}_{\mathrm{bf}}$ is the MAP estimate of the nuisance parameters for the fit to the power spectrum of a \emph{single} HOD model, chosen to be the one that represents the best match to the DESI clustering (e.g. model A0 for LRGs). This estimate of the best-fit nuisance parameters is fixed across the combinations of HOD models computed in \cref{eq:residuals}. The values $\Vec{\Omega}_\mathrm{X}$ and $\Vec{n}_{\mathrm{bf}}$ were determined using the analytic covariance corresponding to the V1 volume for each model described in \cref{sec:cov_mat} with both cosmological and nuisance parameters freely varied. This alternative method removes the contributions to $\vec{\Delta P}^{\mathrm{A,B}}_i(k)$ that are highly correlated in $k$ and are effectively absorbed by the nuisance parameters in any fit, thus isolating only the effects of the HOD variation leading to shifts in the cosmologically-relevant parameters. This leads to a more diagonal $\tens{C}^{\rm V1}_{\rm HOD}$ with greatly reduced amplitude and ensures the total covariance, $\tens{C}^{\rm V1}_{\rm tot}$, is now dominated by the usual statistical term and less sensitive to the precision of the estimated HOD contribution. \cref{fig:shifts} shows the measured cosmological parameter MAP values across the mocks for the entire HOD ensemble. The mean and standard deviation of the fits to 25 individual mock realisations using the V1 volume covariances and `uninformative' nuisance priors are shown in the filled triangles. Fits to the DR1-like power spectrum with covariances corresponding to the DR1 volume, discussed in \cref{sec:cut-sky,sec:cov_mat}, using the baseline `physical' priors, are displayed as crosses. For variations in the parameter values used to compute \cref{eq:residuals}, we are only interested the `V1 MAP (uninformative)' case---fits to the cubic box with the flat nuisance parameters. The effect of the `physical' priors will be added later as an extra contribution. The covariance of these residuals is computed as before, according to \cref{eq:C_HOD}. However, since the determination of the MAP values and the evaluation of $P_{i}(\Vec{\Omega}_\mathrm{X},\Vec{n}_{\mathrm{bf}})$ is done within the context of a cosmological model (in our case, flat \LCDM\ with fixed neutrino mass sum $\sum m_\nu=0.06$ eV), the result is not as general as in the first approach and is instead referred to as the `restricted' approach. In light of this, we also investigate the effect in the $w$CDM model in \cref{sec:wCDM} and find the HOD-dependent systematic contribution to be negligible given the increased statistical errors in this model.

\begin{table}[]
    \centering
    \begin{tabular}{|c|c|c|c|c|}
        \hline
        & \multicolumn{4}{c|}{std($x_\mathrm{A, i} - x_\mathrm{B, i}$) (\% \sigYone)}\\ \cline{2-5}%\hline
        Tracer & $\omega_\mathrm{cdm}$   &  $h$  &  ln($10^{10}A_s$)  &  $n_s$\\ \hline
        BGS & 8.4 & 6.3 & 7.5 & 7.5\\
        LRG & 20.8 & 18.3 & 22.7 & 12.0\\
        ELG & 31.3 & 20.7 & 26.8 & 26.0\\
        QSO & 42.5 & 46.1 & 63.8 & 7.8\\
        \hline
    \end{tabular}
    \caption{Standard deviation of parameter-level shifts in MAP (uninformative) values between V1 cubic HOD mocks. The shifts are quoted relative to the DR1 statistical error (including \texttt{EZmock} rescaling) at the redshift of the data.}
    \label{tab:shifts}
\end{table}

In \cref{sec:full_cov}, we compare the two approaches, `general' and `restricted', in terms of their effects on the final posteriors on cosmological parameters of interest and show that they produce nearly identical results. We present our default results using the `restricted' approach because it leads to a more diagonal covariance contribution and it allows all of the ELG mocks to be incorporated (both those generated at $z=0.8$ and $z=1.1$). However, for future DESI analyses this choice may be revisited.

Computing the covariance of shifts in the power spectrum over all permutations of models A and B (\cref{eq:residuals_true,eq:residuals}) is useful to eliminate sample variance in the halo catalogues. However, the covariance of the shifts is not necessarily a true estimate of the covariance of the power spectrum across the models  (i.e. $\sigma^2(A-B)\neq\sigma^2(A)$). If the models are uncorrelated, the shifts actually lead to an inflation of the estimate standard deviation of the HOD uncertainty by $\sqrt{2}$. Assuming that the HOD models sample some underlying distribution with fixed variance (i.e. $\sigma^2(A)=\sigma^2(B)$), the Cauchy-Schwarz inequality asserts that the combined variance, $\sigma^2(A-B)$, must lie somewhere in the range $0<\sigma^2(A-B)<4\sigma^2(A)$ depending on the level of correlation between models. However, to ensure that our method does not underestimate the systematic contribution, we have verified that
\begin{equation}
    \mathrm{Var}\big[ \vec{P}_{i}(\Vec{\Omega}_\mathrm{A},\Vec{n}_{\mathrm{bf}}) - \vec{P}_{i}(\Vec{\Omega}_\mathrm{B},\Vec{n}_{\mathrm{bf}}) \big] \geq \mathrm{Var}\big[ \vec{P}_{i}(\Vec{\Omega}_\mathrm{A},\Vec{n}_{\mathrm{bf}}) 
    % - \vec{P}_{i}(\Vec{\Omega}_{\mathrm{bf}},\Vec{n}_{\mathrm{bf}})
    \big]
    \label{eq:bound}
\end{equation}
holds across the majority of the $k$-range of interest. The largest violation occurs at high $k$ in the ELG monopole where the variance appears to be underestimated by a factor of $\sim0.8$. This ensures that the systematic estimate presented is conservative, providing an upper bound on the HOD systematic contribution.

Due to the low number of available mocks for BGS and QSO, these mocks are combined with those of the LRG HOD models to produce a more accurate covariance. This is well-motivated given the similarity in the form of their HODs. At the step of creating theory residuals for BGS and QSO following \cref{eq:residuals_true}, supplementary theory power spectra are computed at the corresponding BGS or QSO redshift but instead using the MAP parameter estimates, $\Vec{\Omega}_\mathrm{X}$ and $\Vec{n}_{\mathrm{bf}}$, measured on the LRG mocks. When iterating over permutations in \cref{eq:permutations}, we ensure that only residuals computed across the same tracer are included (i.e. HOD models A and B do not belong to different tracers) as the models have been tuned to a different clustering measurement for each tracer.

The modelling systematic has been shown to be negligible for a $(2\Gpch)^3$ cubic mock populated with a single, fixed HOD \cite{KP5s1-Maus}; however, the DR1 data samples have smaller effective volume and thus lower statistical power than these boxes, so prior choices may have a greater impact. In order to achieve the second key aim of this work, quantifying the influence of nuisance parameter priors, we also include a contribution to the covariance that captures the amplitude of this effect in the DR1 analysis. Given the constraining power of DR1, physically-motivated priors are imposed on nuisance parameters to mitigate projection effects (see \cite{DESI2024.V.KP5}). The physically-motivated stochastic term priors are dependent on the tracer density. Differences in the sample variance and tracer density in DR1 compared to the cubic boxes will change the weight of the prior relative to the likelihood and may systematically shift the MAP value. This shifting of the MAP value due to the miscentering of priors, referred to as the `prior weight effect', is mildly HOD-dependent due to the different values of nuisance parameters required to marginalise over HOD effects as shown in \cref{fig:shifts}. This necessitates the need for analyses with realistic DR1 number density and covariance. To capture all of these effects, we compute an additional diagonal contribution to the covariance
 \begin{equation}
     \vec{D}_\mathrm{model} = \mathrm{max}\big(\big\{ \vec{\delta P}^\mathrm{A} \big\})^2,
     \label{eq:diagonal}
 \end{equation}
from our DR1-like data where
\begin{equation}
    \vec{\delta P}^\mathrm{A} \equiv \vec{\hat{P}}^\mathrm{A}_\mathrm{DR1} - \vec{P}^\mathrm{A}_\mathrm{MAP}.
    \label{eq:mod_res}
\end{equation}
Due to the small number of DR1-like measurements we are able to produce, it is difficult to accurately estimate off-diagonal prior weight effect contributions. For this reason, $\vec{D}_\mathrm{model}$ is constructed as a purely diagonal contribution. The maximum residual in \cref{eq:diagonal} between the DR1-like window-convolved power spectrum (described in \cref{sec:cut-sky}), $\vec{\hat{P}}^\mathrm{A}_\mathrm{DR1}$, and the best-fit window-convolved theory model (evaluated at the MAP estimate of a fit to $\vec{\hat{P}}^\mathrm{A}_\mathrm{DR1}$), $\vec{P}^\mathrm{A}_\mathrm{MAP}$, is computed at each value of $k$ over all HOD models, $\{\mathrm{A}\}$. The MAP fit is performed using the DR1 covariance described in \cref{sec:cov_mat}. In the case of the BGS, $\vec{D}_\mathrm{model}$ cannot practically be computed as we only have access to a single realisation, so we instead assume that it is equal to the equivalent contribution estimated for the LRGs.
 
The total HOD-dependent contribution to the covariance is therefore given by 
\begin{equation}
    \tens{C}_{\rm sys} = \tens{C}^{\rm DR1}_\mathrm{HOD} + \mathrm{diag}(\vec{D}_\mathrm{model}).
\end{equation}

\section{Results}
\label{sec:results}

\begin{figure}
    \centering
    \includegraphics[width=\linewidth]{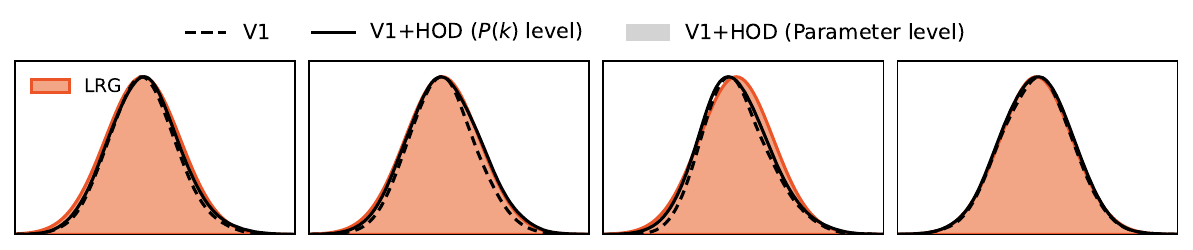}
    \includegraphics[width=\linewidth]{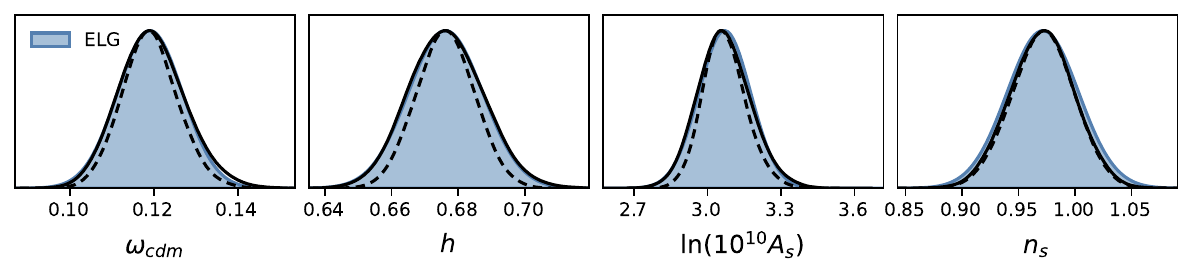}
    \caption{Parameter posteriors when HOD systematics are added at the parameter-level (\emph{filled}) versus at the data-level (\emph{solid}) for LRG (\emph{red}) and ELG (\emph{blue}) tracer samples. Fits are performed to the V1 cubic mock populated with the best-fit HOD model for each tracer. The V1-only posterior with no HOD systematic contribution is given as a dashed line for comparison. The relative HOD contribution is significantly greater in the case of V1 than compared to DR1 data due to the large increase in volume. The posterior on $n_s$ is prior-dominated and we therefore do not expect to see any change with the HOD contribution included at the $P(k)$-level.}
    \label{fig:validation}
\end{figure}

\subsection{Comparison to parameter-level estimates}
\label{sec:shifts}

In this section, we compare the methods of adding a HOD-dependent systematic contribution at the level of the parameters---computed as described below---and at the level of the power spectrum.  \cref{fig:validation} shows resulting parameter distributions for the combination of the statistical uncertainty, given the V1 cubic box volume, and the HOD-dependent systematic contribution, which we refer to as `V1+HOD'. The two approaches are compared in the context of the V1 cubic box volume rather than the reduced volume of DR1 in order to more easily distinguish the HOD contribution from the statistical error. To show the combined V1+HOD uncertainty at the parameter-level, we generate a Gaussian distribution centred at the posterior mean of a fit to the mean power spectrum obtained from 25 realisations of the V1 cubic mock. The posterior width, $\sigma^\mathrm{V1}_\mathrm{stat}$, on a given cosmological parameter of interest, $x$, is then inflated by the HOD contribution in quadrature,

\begin{equation}
    (\sigma^{x}_\mathrm{comb})^2 = (\sigma^\mathrm{V1}_\mathrm{stat})^2 + (\sigma^\mathrm{V1}_\mathrm{HOD})^2 + \mathrm{max}\big(
    \{\bar{x}^\mathrm{A}_p - \bar{x}^\mathrm{A}_\mathrm{flat}\}
    \big)^2
    \label{eq:parameter_level}
\end{equation}
where $\bar{x}_p$ and $\bar{x}_\mathrm{flat}$ are the MAP values fit to the mean of 25 mocks with physical and uninformative nuisance priors, respectively.
The final term captures the prior weight effect (given the V1 volume) in a similar manner to \cref{eq:diagonal} while the contribution from HOD-dependent shifts in cosmology, 
\begin{equation}
    \sigma^\mathrm{V1}_\mathrm{HOD} \equiv \mathrm{std}\big(\{ x^\mathrm{A}_{i} - x^\mathrm{B}_{i} \}_{\mathrm{A}\neq\mathrm{B}}\big),
    \label{eq:parameter_shifts}
\end{equation}
is the parameter-level equivalent to \cref{eq:C_HOD}, with values provided in \cref{tab:shifts} for each tracer. Here, $x^\mathrm{X}_{i}$ denotes the MAP values of HOD model X fit to mock realisation $i$ with uninformative nuisance priors. In this comparison, the V1 analytic covariances described in \cref{sec:cov_mat} were used throughout. The posterior resulting from our fiducial analysis with the HOD contribution added at the level of the power spectrum, as described in the previous section, is shown for comparison. To produce this posterior for the V1 volume, the analytic covariance was combined with the unwindowed HOD contribution (\cref{eq:C_HOD}) including a diagonal contribution equivalent to that of \cref{eq:diagonal}, except that the residual shown in \cref{eq:mod_res} between best-fit model and data was determined from a fit to the mean power spectrum obtained from 25 cubic box mock realisations using the V1 analytic covariance rather than the DR1-like data vectors. The figure shows exceptional agreement between the two cases demonstrating that both the inability of the model to absorb changes in the galaxy-halo connection and the impact of the priors on nuisance parameters are correctly accounted for. The additional diagonal contribution to the HOD covariance captures the effect of the nuisance priors in a trivial way---shifts in the MAP estimate are simply translated into the ability of the model to fit the data with the chosen priors.

\subsection{DR1 HOD covariance}

\begin{figure}
    \centering
    \includegraphics[width=\linewidth]{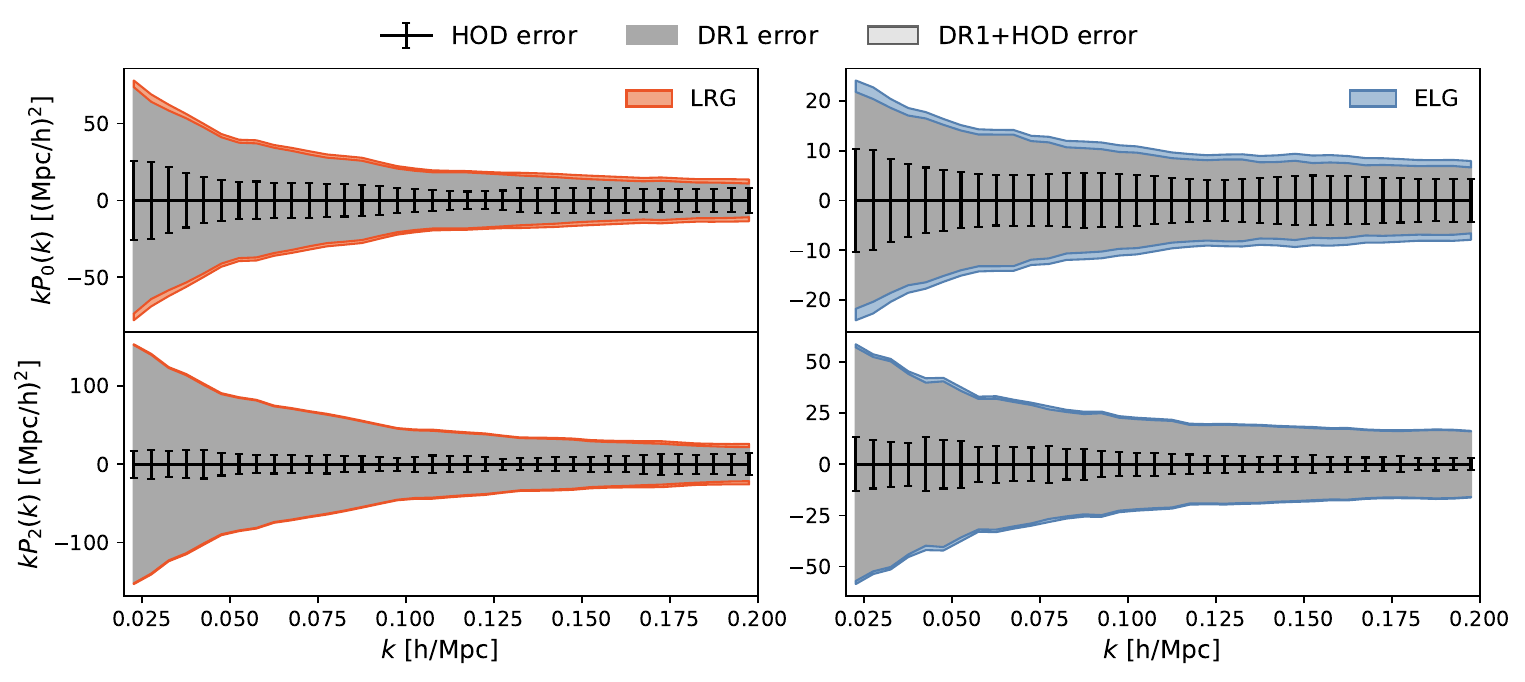}
    \caption{Contribution of the estimated HOD-dependent systematic uncertainty to the diagonal of the DR1 \texttt{EZmock} covariance matrix.}
    \label{fig:diagonal}
\end{figure}

\begin{figure}
    \centering
      \includegraphics[width=\linewidth]{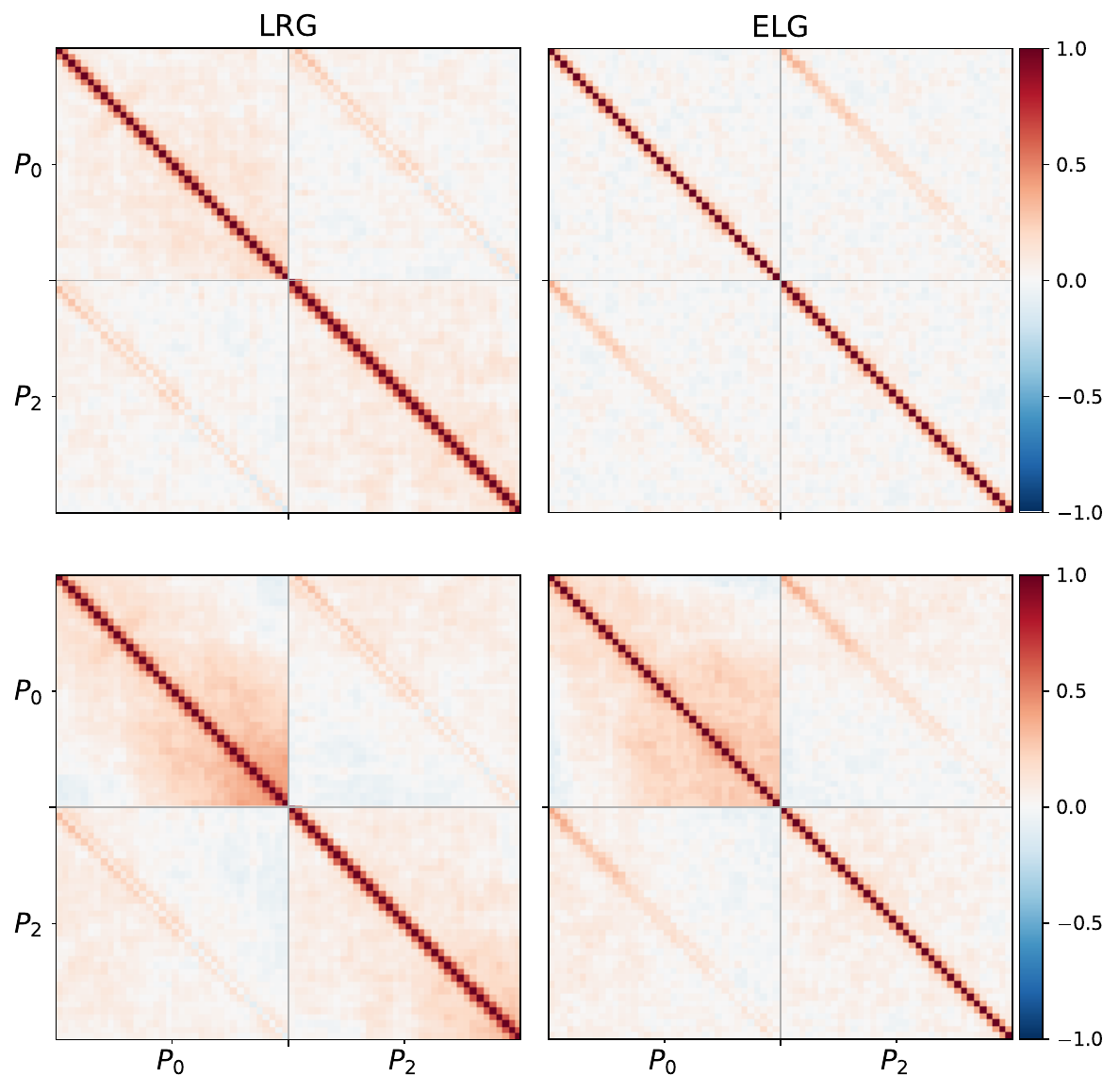}
    \caption{Correlation matrix for DR1 estimated from 1000 \texttt{EZmocks} (\emph{upper}) and DR1+HOD combined (\emph{lower}). Off-diagonal correlations are a result of the HOD-dependent contribution.}
    \label{fig:cov_mat}
\end{figure}

With our method of treating the systematics validated on the cubic mocks, we present results for the final DR1 windowed covariance. \cref{fig:diagonal} shows the additional contribution to the \texttt{EZmock} covariance diagonal. The \texttt{EZmock} rescaling, discussed in \cref{sec:cov_mat}, has not been applied in the figure to make the HOD contribution more apparent. As expected, the uncertainty sourced by varying the galaxy-halo connection is most dominant at small scales relative to the statistical uncertainty. However, it is also evident that even the largest scales are impacted by the inability to completely marginalise over these small-scale effects. The full combined correlation matrices are shown in \cref{fig:cov_mat}. The HOD contribution has a higher degree of correlation than the \texttt{EZmock} statistical covariance but this off-diagonal contribution is subdominant to the diagonal of the statistical covariance as a result of the `restricted' method and has no effect on the parameter correlation structure (see \cref{fig:corner} in \cref{sec:fits}).

\subsection{Combined covariance fits to DR1 mocks}
\label{sec:fits}

\begin{figure}
    \centering
    \includegraphics[width=0.8\linewidth]{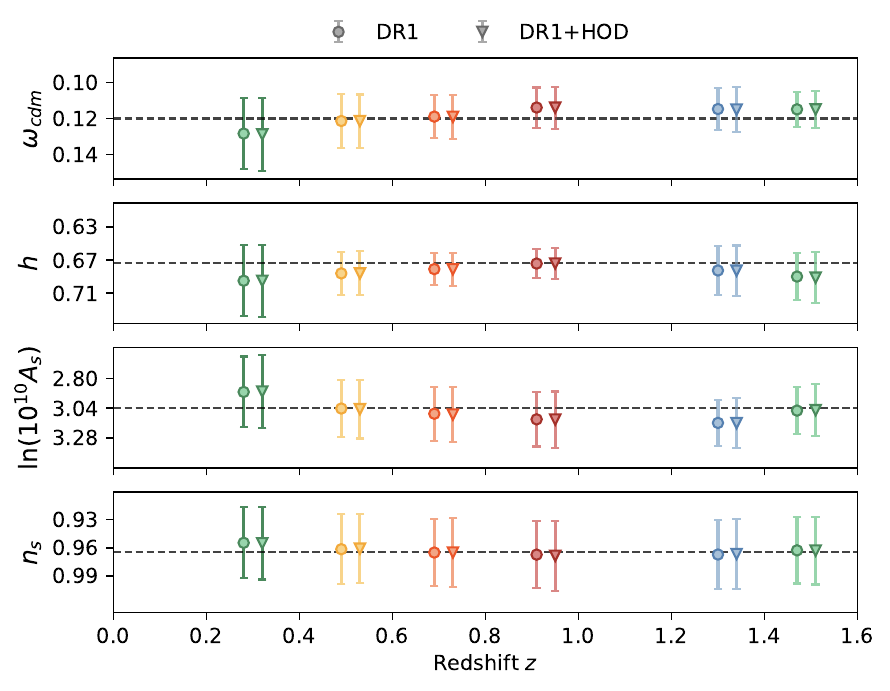}
    \caption{Effect of the HOD systematic contribution on fits to DR1 mocks. Mean and $1\sigma$ intervals are shown for each tracer and centred at the redshift of the data for visualisation purposes. The DR1 covariance is computed from 1000 \texttt{EZmocks}.}
    \label{fig:tracer_fit}
\end{figure}

The DESI 2024 Full-Shape analysis utilises the covariance matrices produced in this work, in combination with \texttt{EZmock}-based covariance matrices, in order to determine the full statistical plus systematic error. Mock-based covariance matrices have an intrinsic uncertainty in their estimate and also result in biased estimates of the inverse \cite{Hartlap2007, Dodelson2013, Percival2014, Sellentin2016}. To account for this, a correction factor, 
\begin{equation}
    f = \frac{(n_\mathrm{s} - 1) \big[1 + B(n_\mathrm{d}-n_\theta)\big]}{
    (n_\mathrm{s} - n_\mathrm{d} + n_\theta - 1)
    }
\end{equation}
with
\begin{equation}
    B = \frac{(n_\mathrm{s} - n_\mathrm{d} - 2)}{(n_\mathrm{s} - n_\mathrm{d} - 1)(n_\mathrm{s} - n_\mathrm{d} - 4)},
\end{equation}
is typically applied, where $n_\mathrm{s}$, $n_\mathrm{d}$ and $n_\theta$ are the number of mock samples, data points and parameters, respectively \cite{Percival2022}. This is a generalisation of the Hartlap correction \cite{Hartlap2007} to also propagate the uncertainty in the estimate of the covariance matrix to the derived parameter posteriors. This factor is used with a Gaussian likelihood form as a good approximation to the correct treatment, which is to modify the likelihood itself \cite{Sellentin2016}. We apply this correction only to the \texttt{EZmock} statistical covariance, $\tens{C}_{\rm stat}$, given that the expression
\begin{align}
    \langle (\tens{C}_{\rm stat} + \tens{C}_{\rm sys})^{-1} \rangle &\approx \langle \tens{C}_{\rm stat}^{-1} \rangle - \langle \tens{C}_{\rm stat}^{-1} \rangle \tens{C}_{\rm sys} \langle \tens{C}_{\rm stat}^{-1} \rangle \label{eq:combined_inverse}\\
    &\approx (f \tens{C}_{\rm stat} + \tens{C}_{\rm sys})^{-1} \label{eq:cov_correct}   
\end{align}
holds to first order under the condition that $\tens{C}_{\rm sys}$ is a small contribution to $\tens{C}_{\rm stat}$. While this assumes that $\tens{C}_{\rm sys}$ is perfectly known, noise in the estimate of $\tens{C}_{\rm sys}$ should be negligible in terms of the total covariance.

\begin{figure}
    \centering
    \includegraphics[width=.8\linewidth]{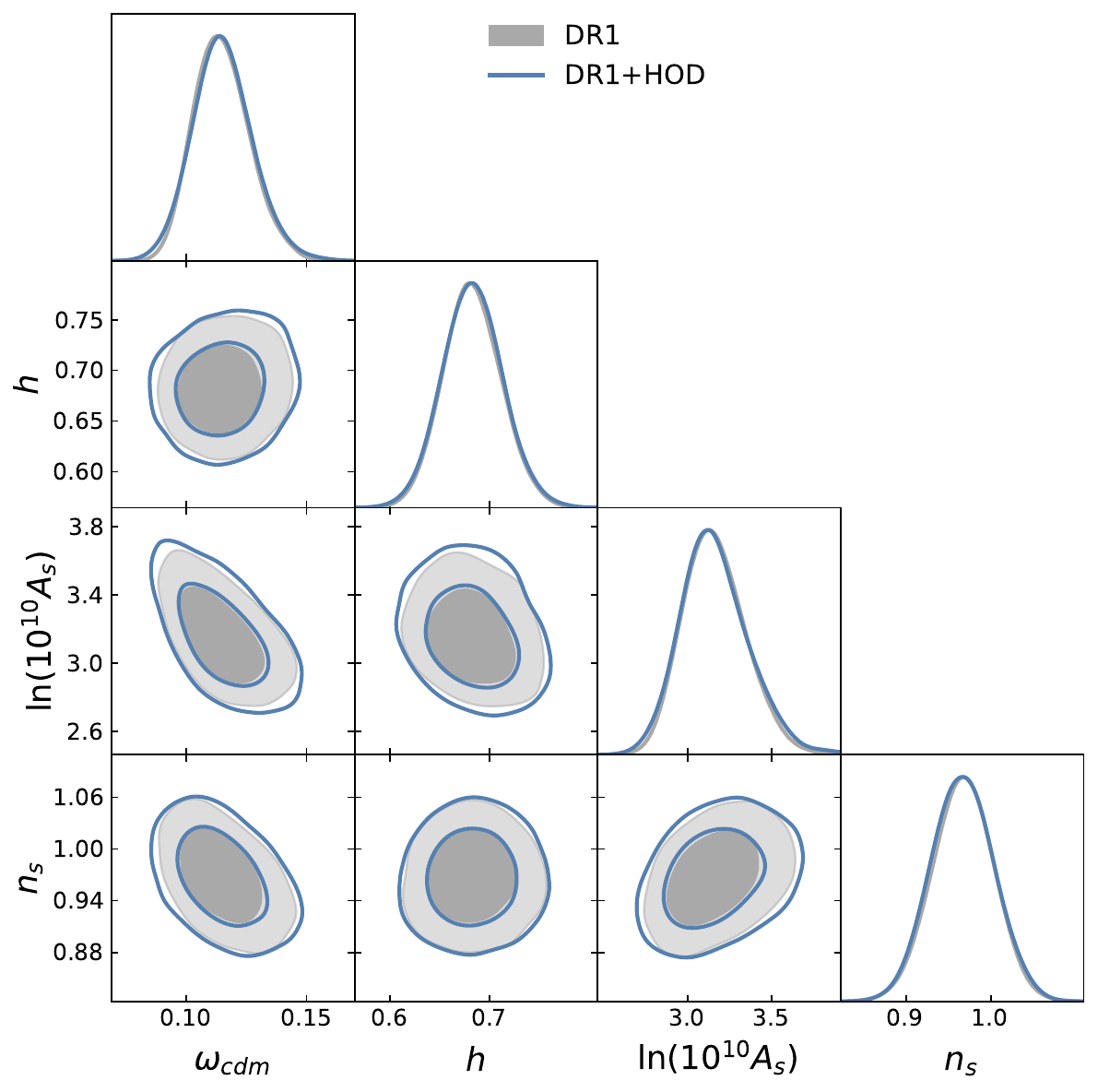}
    \caption{Cosmological parameter posteriors for the ELG DR1 mock at $z=1.1-1.6$. By utilising a full covariance treatment, our method of adding the HOD systematic contribution at the level of the data vector allows the effect on the 2D posterior to be shown. The effect on the posterior is minimal.}
    \label{fig:corner}
\end{figure}

\cref{fig:tracer_fit} shows that the effect of the additional HOD systematic covariance is minimal for fits to DR1 mocks. However, this will become more prevalent as the constraining power of the survey increases, becoming a significant contribution to the total error budget for a V1-like volume (see \cref{fig:validation}). The mean values in \cref{fig:tracer_fit} have been plotted at the effective redshifts of the data for visualisation purposes. As we employ a full covariance treatment, the effect on the 2-dimensional posterior can also be explored in \cref{fig:corner}. For the ELG tracer in the redshift range $z=1.1-1.6$, we show that the HOD contribution acts to inflate the cosmological parameter contours while maintaining the degeneracy structure. For brevity, we do not show the equivalent LRG posterior as the difference after adding the HOD contribution is even less pronounced.

\section{Conclusions}
\label{sec:conclusions}

In this paper, we have studied the impact of varying the HOD on the DESI 2024 Full-Shape galaxy clustering analysis, and presented a new method for the inclusion of mock-based systematic estimates at the level of the data vector. By fitting an EFT model to a variety of HOD mocks for the four DESI tracers---BGS, LRG, ELG and QSO, we have produced systematic covariance matrices that reflect the HOD-dependent variation of the data vector. Additionally, our systematic covariance includes a contribution that captures the ability of the model to fit the HOD mocks given a set of informative nuisance parameter priors---naturally incorporating any uncertainty in our choice of prior. Our method has been validated against the parameter-level approach used formerly \cite{KP4s10-Mena-Fernandez,KP4s11-Garcia-Quintero}, showing excellent consistency. The HOD systematic covariance matrices for each tracer are provided for the DESI 2024 Full-Shape analysis as an additional contribution to the statistical covariance.

At the parameter-level, changes in the HOD have been shown to shift the recovered cosmological parameters by greater than 20\% of the DR1 statistical error (\cref{tab:shifts}). This only introduces a near-negligible inflation of the posterior width for each of the samples. However, we do expect this effect to become more important as the constraining power of the survey improves, as evidenced by the effect on the $(2\Gpch)^3$ V1 cubic mocks (\cref{fig:validation}). 

Adding the systematic contribution at the level of the data vector has advantages. A full covariance treatment is more rigorous than simply inflating the statistical covariance by some factor or broadening uncertainties on the recovered parameters. The method should also be more general and robust to the addition of external datasets or choice of model. The covariance matrices provided for the DESI 2024 Full-Shape analysis employ a method that loses some generality to modelling choices in order to diagonalise and reduce sensitivity to the covariance (\cref{eq:residuals}). This `restricted' method yields a more conservative estimate in \LCDM\ but is, in general, not fully applicable to other cosmologies given that these often introduce new degeneracies between cosmology and nuisance parameters. However, in \cref{sec:wCDM}, we show that the choice of method is arbitrary for extended models in DR1 given the size of statistical uncertainties. While we take the conservative approach and employ the `restricted' method for DR1, we may reconsider this choice in future data releases as the constraining power of the survey increases. 

Incorporating the HOD-dependent systematic at the level of the data vector is a method that could be applied to other mock-based systematics tests provided a suitably large number of mocks are available. Given that this method is not suited for systematic tests with a low number of mocks, increasing the number of HOD mocks for the BGS and QSO samples is of high priority. Additionally, exploring the effects of varying the HOD in mocks created with non-\LCDM\ base cosmologies is an essential step forward in light of results from DESI BAO \cite{DESI2024.VI.KP7A}. While we expect the cosmological dependence of the HOD covariance to be small, our generalised method can be easily extended to include alternate cosmologies provided that a sufficient number of mocks are available.

\section*{Data Availability}

Data from the plots in this paper will be available on Zenodo as part of DESI's Data Management Plan.
The data used in this analysis will be made public along with Data Release 1 (details in \href{https://data.desi.lbl.gov/doc/releases/}{https://data.desi.lbl.gov/doc/releases/}).

\acknowledgments
We would like to acknowledge Mark Maus and Kazuya Koyama for serving as internal reviewers of this work and providing useful feedback. We thank Samuel Brieden for a comment on the limitation of fixing nuisance parameters in the creation of the HOD covariance that helped to shape this paper. NF acknowledges support from STFC grant ST/X508688/1 and funding from the University of Portsmouth. SN acknowledges support from an STFC Ernest Rutherford Fellowship, with grant reference ST/T005009/2. CGQ acknowledges support provided by NASA through the NASA Hubble Fellowship grant HST-HF2-51554.001-A awarded by the Space Telescope Science Institute, which is operated by the Association of Universities for Research in Astronomy, Inc., for NASA, under contract NAS5-26555.

This material is based upon work supported by the U.S. Department of Energy (DOE), Office of Science, Office of High-Energy Physics, under Contract No. DE–AC02–05CH11231, and by the National Energy Research Scientific Computing Center, a DOE Office of Science User Facility under the same contract. Additional support for DESI was provided by the U.S. National Science Foundation (NSF), Division of Astronomical Sciences under Contract No. AST-0950945 to the NSF National Optical-Infrared Astronomy Research Laboratory; the Science and Technology Facilities Council of the United Kingdom; the Gordon and Betty Moore Foundation; the Heising-Simons Foundation; the French Alternative Energies and Atomic Energy Commission (CEA); the National Council of Humanities, Science and Technology of Mexico (CONAHCYT); the Ministry of Science and Innovation of Spain (MICINN), and by the DESI Member Institutions: \url{https://www.desi.lbl.gov/collaborating-institutions}. Any opinions, findings, and conclusions or recommendations expressed in this material are those of the author(s) and do not necessarily reflect the views of the U. S. National Science Foundation, the U. S. Department of Energy, or any of the listed funding agencies.

The authors are honored to be permitted to conduct scientific research on Iolkam Du’ag (Kitt Peak), a mountain with particular significance to the Tohono O’odham Nation.

\appendix

\section{Shot-noise contribution to estimates of the HOD-dependent uncertainty}
\label{sec:shotnoise}

\begin{figure}
    \centering
    \includegraphics[width=.8\linewidth]{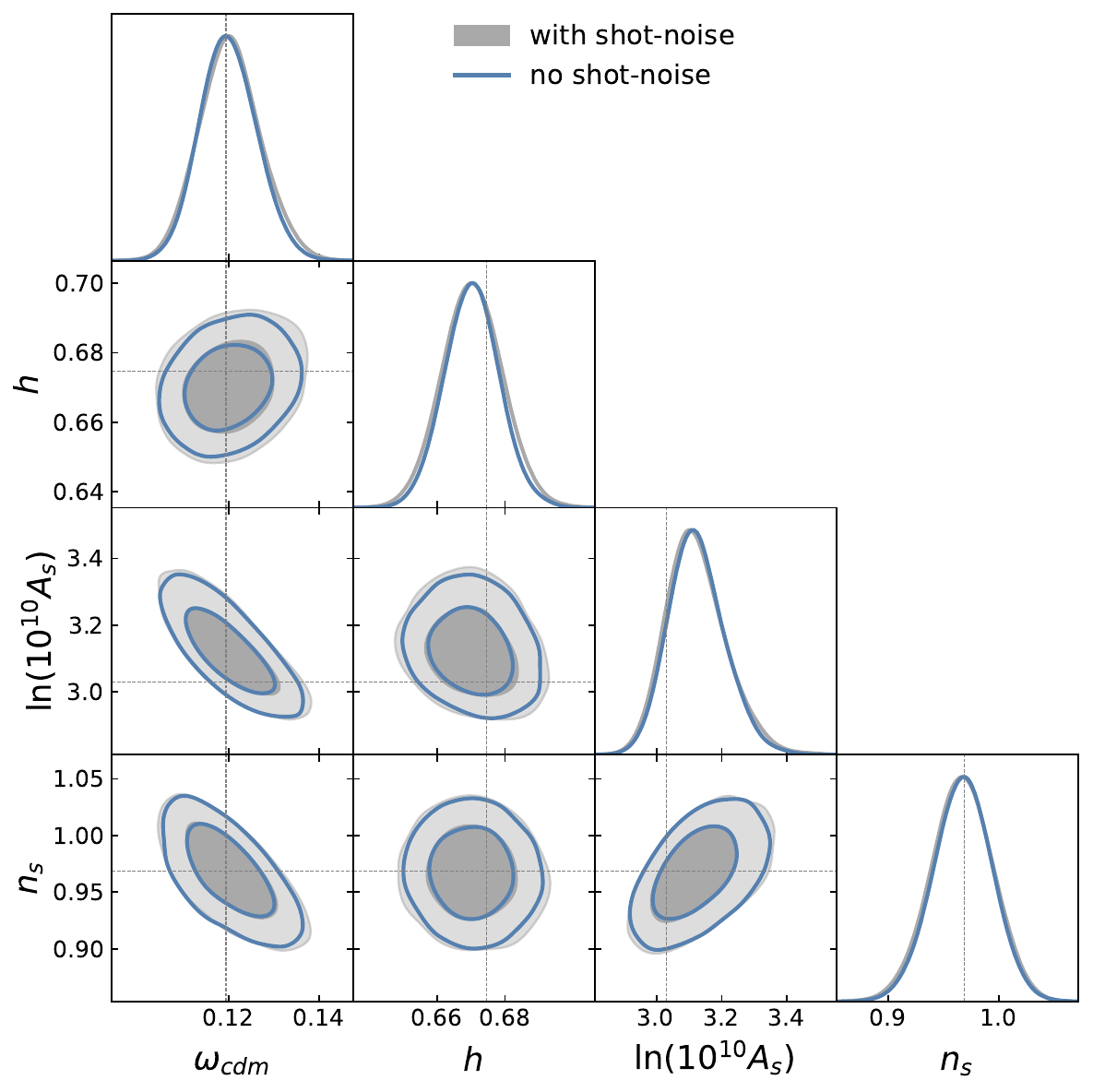}
    \caption{Cosmological parameter posteriors for synthetic ELG data with added noise generated using analytic covariance matrices described in \cref{sec:cov_mat}. The covariances correspond to a volume of $(2\Gpch)^3$ equivalent to that of the cubic \Abacus mocks and have been generated with and without shot-noise contributions. The shot-noise contribution is on the order of a few percent.}
    \label{fig:shotnoise}
\end{figure}

In a given mock power spectrum measurement, contributions to uncertainty arise from a combination of the particular realisation of the density field (i.e. the initial conditions), the underlying halo catalogue and the stochastic process of populating those halos with galaxies. Estimates of the HOD contribution using either \cref{eq:residuals_true} or \cref{eq:residuals} eliminate both noise from the initial conditions and the halo catalogue by fixing the mock realisation when computing power spectrum shifts. However, noise contributions attributed to the galaxy catalogue remain. This has two components: (i) uncertainty in the galaxy-halo connection itself (i.e. the HOD model according to which halos are assigned galaxies), and (ii) shot-noise from the finite number of galaxies that sample the underlying HOD model. The latter is not of interest for this work, given that it is included in the statistical covariance. The former, however, contributes to our HOD-dependent systematic contribution and is present even in the case of infinite tracer density. Comparing the differences between the galaxy-halo models in individual realisations while holding fixed the initial conditions and the halo catalogue, as we do in determining the mock-to-mock differences, helps to isolate the contribution of the uncertainty in the galaxy-halo connection, which can otherwise be washed out when averaging each model over many mock realisations before comparing differences. However, there is still a shot-noise contribution associated with sampling the halo catalogue with a finite number of galaxies which may also affect estimates derived this way. Our goal here is to demonstrate that the total shot-noise effect on the recovered parameter uncertainties is small and that this contribution can therefore be neglected when estimating the HOD contribution to the systematic error budget.

In order to do this, we investigate the level of shot-noise one would expect in a measurement of the cosmological parameters using a single mock realisation. The ELG tracer was selected for this investigation as it displays one of the largest estimated HOD-uncertainty contributions (see \cref{tab:shifts}). To achieve this, two analytic ELG covariance matrices were produced with a volume corresponding to the $(2\Gpch)^3$ cubic \Abacus mock following the method in \cref{sec:cov_mat}. The first was computed with Poissonian shot-noise corresponding to that of the ELG cubic mock (with galaxy density $\bar{n}\approx 2.4\times10^{-3} \hhhMpc$) and the second with zero shot-noise contribution ($\bar{n}\to\infty$). These covariances were then sampled in order to generate synthetic noise that could be added to a noiseless theoretical power spectrum $\Vec{P}_\mathrm{th}$ following
\begin{equation}
    \label{eq:Cholesky}
    \Vec{P}_\mathrm{noisy} = \Vec{P}_\mathrm{th} + \tens{L}\Vec{Z},
\end{equation}
where $\tens{L}$ is the Cholesky decomposition of the covariance matrices described above and $\Vec{Z}$ is a vector of independent standard normal random variables. The noisy data vectors were then sampled using their corresponding covariance to obtain the posterior distribution of our cosmological parameters of interest. \cref{fig:shotnoise} shows the resulting posteriors for the cases with and without the shot-noise contribution.

The width of these posteriors can be compared to the parameter shifts measured in \cref{tab:shifts}. Given that the total variance of parameter $x$,
\begin{equation}
    \sigma_\mathrm{x, \,tot}^2 = \sigma_\mathrm{x,\, SV}^2 + \sigma_\mathrm{x,\, SN}^2,
\end{equation}
is composed of a sample variance part, $\sigma_\mathrm{x,\,SV}^2$, and a shot-noise part, $\sigma_\mathrm{x,\,SN}^2$, we can isolate the shot-noise contribution directly from the standard deviations of the marginalised posteriors. This leads to the expression
\begin{equation}
    \frac{\sigma_\mathrm{\Delta x,\, SN}}{\sigma^\mathrm{DR1}_\mathrm{stat}} = \frac{\sqrt{2(\sigma_\mathrm{x, \,tot}^2 - \sigma_\mathrm{x,\, SV}^2)}}{\sigma^\mathrm{DR1}_\mathrm{stat}}
    \label{eq:shotnoise}
\end{equation}
for the shot-noise contribution, $\sigma_\mathrm{\Delta x,\, SN}$, to the measured parameter shifts relative to DR1 statistical uncertainties, $\sigma^\mathrm{DR1}_\mathrm{stat}$. The factor of $\sqrt{2}$ must be included in order to correctly estimate the standard deviation of \emph{shifts} in the parameter and not only the standard deviation of the parameter itself. Using \cref{eq:shotnoise}, the contribution of shot-noise was determined to be on the order of a few percent, and hence is negligible compared to the HOD-dependent estimates listed in \cref{tab:shifts}.

\section{Method consistency in \LCDM}
\label{sec:full_cov}

\begin{figure}
    \centering
    \includegraphics[width=\linewidth]{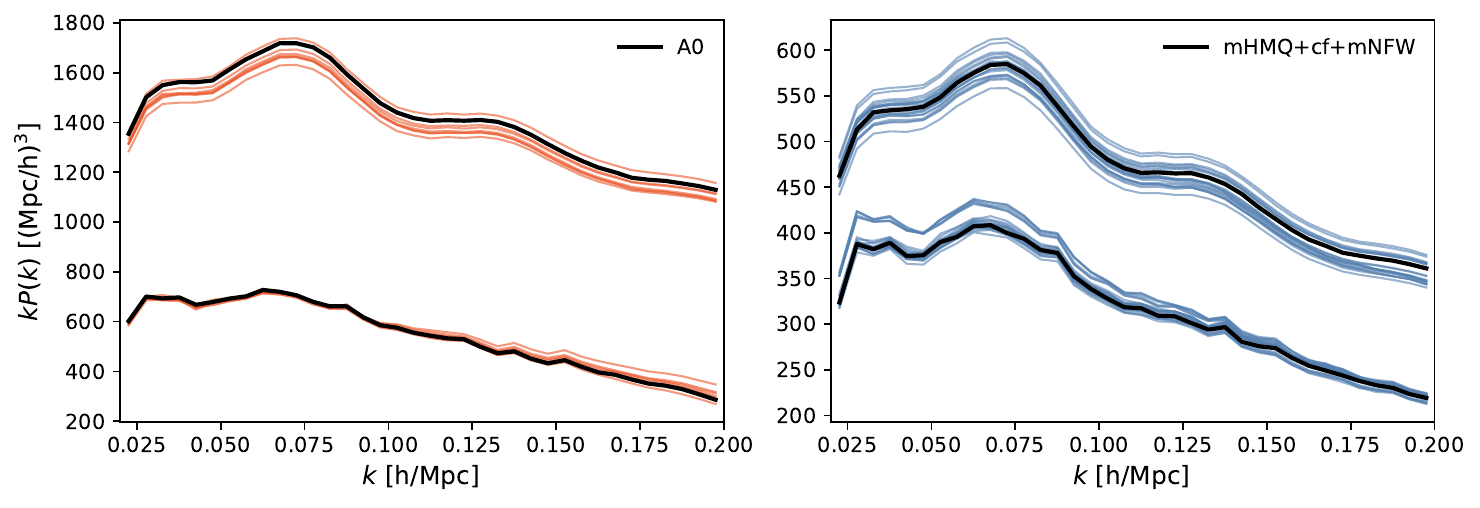}
    \caption{Monopole and quadrupole power spectrum measurements of HOD-varied \Abacus cubic mocks (\cref{sec:pks}) averaged over 25 realisations for the LRG (\emph{left}) and ELG (\emph{right}) samples. Each line displays a different model of the HOD ensemble we explore, with the model that provides the best fit to the DESI One-Percent data highlighted in black.}
    \label{fig:HODs}
\end{figure}

\begin{figure}
    \centering
    \includegraphics[width=\linewidth]{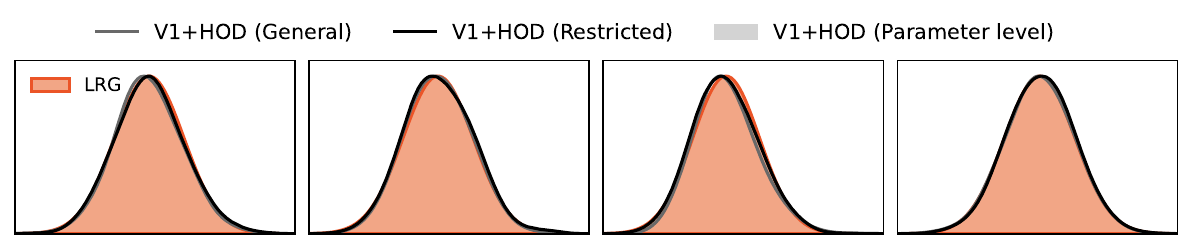}
    \includegraphics[width=\linewidth]{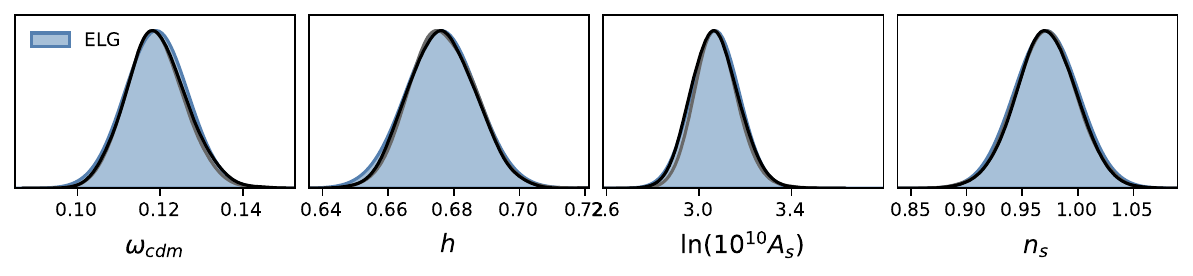}
    \caption{Comparison of the effect of using the HOD covariance matrix generated using the generalised approach (\emph{grey}) and the `restricted' approach in which nuisance parameters are fixed (\emph{black}). ELG models at $z=0.8$ and the diagonal contribution of \cref{eq:diagonal} have not been included here for consistency. Both methods at the level of the data vector are able to reproduce the parameter-level HOD contribution of \cref{eq:parameter_level} without the diagonal contribution (\emph{filled}).}
    \label{fig:fixed_nuisance}
\end{figure}

In this work, we choose to take a `restricted' approach in the computation of the covariance matrix in order to produce a more diagonal and less sensitive covariance. In this approach, the nuisance parameters are fixed to the measured best-fit values corresponding to a single HOD model following \cref{eq:residuals}. Alternatively, when the covariance is computed directly from the measured power spectra following the `general' method of \cref{eq:residuals_true}, the covariance is large in amplitude and highly non-diagonal due to the different effective galaxy biases of the HOD mocks. These galaxy biases lead to highly correlated shifts in the mock power spectra, shown in \cref{fig:HODs}, that are a result of fitting to the small-scale clustering only and thereby leaving large-scale effects unconstrained. These shifts are absorbed by the nuisance parameters of the EFT model therefore removing this correlation from the `restricted' covariance.

The generalised approach, following \cref{eq:residuals_true}, produces a highly-correlated covariance matrix with a magnitude of the order of the statistical covariance. Given the large relative contribution to the total covariance, greater accuracy in estimating the correct correlation structure is required. One must also take care in applying covariance correction factors (see \cref{sec:fits}) to the, now non-negligible, HOD contribution. As we are using analytically-determined statistical covariance matrices, we instead apply correction factor $f$ to the HOD contribution only. This follows the same line of thought as \cref{eq:combined_inverse} but instead treats $\tens{C}_{\rm stat}$ as perfectly known and accounts for noise in the estimate of $\tens{C}_{\rm HOD}$. \cref{fig:fixed_nuisance} shows that the number of mocks used in this work is sufficient to achieve equivalent posteriors with these approaches in \LCDM.

Although not fully generalisable to other cosmologies (see \cref{sec:wCDM}), the `restricted' method should remain far more general than simply transferring the uncertainty determined using parameter shifts measured with one dataset combination (e.g. Full-Shape alone) to another (e.g. BAO + Full-Shape). Both methods presented in this paper add the Full-Shape alone HOD-uncertainty to the power spectrum covariance such that, when used in combination with other datasets, the relative uncertainty contribution will be correctly accounted for in the likelihood. Additionally, generating the covariance using the `restricted' approach also allows the inclusion of HOD mocks created at a different redshift as the theory predictions used to compute \cref{eq:residuals} can be evaluated at any redshift with the same best-fit mock cosmologies. This is advantageous for the DR1 analysis given that the ELG models are split over two redshift bins. Although we proceed with the `restricted' method for the DR1 analysis for robustness, \cref{fig:fixed_nuisance} highlights that both methods are entirely consistent within \LCDM\ and motivate the full, general approach for future data releases.

\section{HOD-dependence and performance of `restricted' method in $w$CDM}
\label{sec:wCDM}

\begin{figure}
    \centering
    \includegraphics[width=\linewidth]{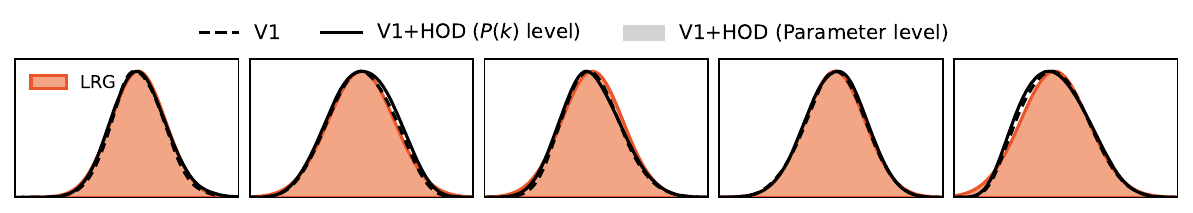}
    \includegraphics[width=\linewidth]{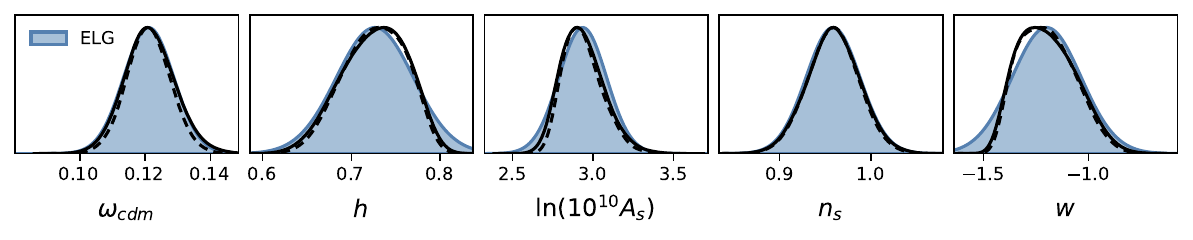}
    \caption{Same as \cref{fig:validation} except the equation of state parameter, $w$, has been varied when estimating the parameter-level contribution and during sampling. We compare the parameter-level contribution estimated in $w$CDM (\emph{filled}) to the `restricted' method at the level of the data-vector estimated in \LCDM\ (\emph{solid}) and find minimal difference. This suggests that, given the small relative HOD-dependent contribution, the `restricted' method estimated in \LCDM\ is sufficient for extended models. The filled contours are Gaussian curves and so do not reflect the non-Gaussianity of the true posteriors given by the solid and dashed lines.}
    
    \label{fig:wCDM_post}
\end{figure}

In order to test the robustness of our method in extended cosmologies, we explore HOD-dependent systematics within the framework of the $w$CDM cosmological model. When the equation of state parameter, $w$, is allowed to vary, we find that the shifts in cosmological parameters between different HOD mocks are larger than in the \LCDM\ case. This is because of the introduction of new degeneracies between $w$ and the other parameters. Immediately this highlights one of the pitfalls of the parameter-level method as it implies that parameter-level estimates measured in one cosmology cannot be transferred to another cosmology. This is due to two reasons: (i) the HOD-dependence of new parameters (i.e. $w$) cannot be estimated in the original cosmology (i.e. \LCDM ) and (ii) new degeneracies introduced by additional parameters will change how the HOD-dependence affects the cosmological parameters. The `general' method proposed in this work should solve both of these problems by quantifying the HOD-dependent variation of the data-vector, making no assumption of model parametrisation. While the `restricted' method should be more robust to these effects than the parameter-level method, it struggles with regards to  nuisance parameter degeneracies.

In the `restricted' method, the nuisance parameters are fixed to a single set of values for every HOD model included in the covariance estimation. Therefore, the uncertainty estimated using this method is only valid in cases where the relationship to the nuisance parameters is unaffected. In extended models such as $w$CDM, this is not the case. The new degeneracy with nuisance parameters leads to variations in the cosmological parameters that are larger than in \LCDM\ (as these are compensated by larger variations in the nuisance parameters). When the nuisance parameters are fixed in the `restricted' method, the variations in cosmology in \LCDM, and hence HOD covariance contribution, are underestimated compared to those you would measure in $w$CDM. This shortcoming in extending to other cosmologies/parametrisations is why the method is referred to as `restricted'. However, \cref{fig:wCDM_post} shows that the difference in the posterior distributions estimated using the correct parameter-level HOD-contribution as measured in $w$CDM compared to using the `restricted' data-vector-level method in \LCDM\ is negligible for the V1 cubic box due to the increased statistical uncertainty in $w$CDM. Given that extended models are unlikely to be significantly impacted by HOD-dependent systematics for DR1 due to the large statistical uncertainty, we motivate using the `restricted' method given that it is more conservative in \LCDM.

\input{HOD_author_list.affiliations}

\bibliographystyle{JHEP}
\bibliography{references, DESI_supporting_papers}{}

\end{document}

%% file: HOD_author_list.tex
\emailAdd{nathan.findlay@port.ac.uk}
\emailAdd{seshadri.nadathur@port.ac.uk}
\affiliation{Affiliations are in Appendix \ref{sec:affiliations}}

\author[1]{{N.~Findlay}\orcidlink{0009-0007-0716-3477},}
\author[1]{{S.~Nadathur}\orcidlink{0000-0001-9070-3102},}
\author[2,3,4]{{W.~J.~Percival}\orcidlink{0000-0002-0644-5727},}
\author[5]{{A.~de~Mattia}\orcidlink{0000-0003-0920-2947},}
\author[6]{{P.~Zarrouk}\orcidlink{0000-0002-7305-9578},}
\author[7,8,9]{{H.~Gil-Mar\'in}\orcidlink{0000-0003-0265-6217},}
\author[10]{{O.~Alves},}
\author[11]{{J.~Mena-Fern\'andez}\orcidlink{0000-0001-9497-7266},}
\author[12,13,14]{{C.~Garcia-Quintero}\orcidlink{0000-0003-1481-4294},}
\author[15,5]{{A.~Rocher}\orcidlink{0000-0003-4349-6424},}
\author[16]{{S.~Ahlen}\orcidlink{0000-0001-6098-7247},}
\author[17]{{D.~Bianchi}\orcidlink{0000-0001-9712-0006},}
\author[18]{{D.~Brooks},}
\author[19]{{T.~Claybaugh},}
\author[20]{{S.~Cole}\orcidlink{0000-0002-5954-7903},}
\author[21]{{A.~de la Macorra}\orcidlink{0000-0002-1769-1640},}
\author[22]{{A.~Dey}\orcidlink{0000-0002-4928-4003},}
\author[18]{{P.~Doel},}
\author[23,24]{{K.~Fanning}\orcidlink{0000-0003-2371-3356},}
\author[18,25]{{A.~Font-Ribera}\orcidlink{0000-0002-3033-7312},}
\author[26,27]{{J.~E.~Forero-Romero}\orcidlink{0000-0002-2890-3725},}
\author[8,1,28]{{E.~Gaztañaga},}
\author[29]{{G.~Gutierrez},}
\author[30]{{C.~Hahn}\orcidlink{0000-0003-1197-0902},}
\author[31,32,33]{{K.~Honscheid},}
\author[34]{{C.~Howlett}\orcidlink{0000-0002-1081-9410},}
\author[22]{{S.~Juneau},}
\author[19]{{M.~E.~Levi}\orcidlink{0000-0003-1887-1018},}
\author[22]{{A.~Meisner}\orcidlink{0000-0002-1125-7384},}
\author[35,25]{{R.~Miquel},}
\author[36]{{J.~Moustakas}\orcidlink{0000-0002-2733-4559},}
\author[5,19]{{N.~Palanque-Delabrouille}\orcidlink{0000-0003-3188-784X},}
\author[37]{{I.~P\'erez-R\`afols}\orcidlink{0000-0001-6979-0125},}
\author[38]{{G.~Rossi},}
\author[39]{{E.~Sanchez}\orcidlink{0000-0002-9646-8198},}
\author[19]{{D.~Schlegel},}
\author[40,10]{{M.~Schubnell},}
\author[41]{{H.~Seo}\orcidlink{0000-0002-6588-3508},}
\author[22]{{D.~Sprayberry},}
\author[10]{{G.~Tarl\'{e}}\orcidlink{0000-0003-1704-0781},}
\author[21]{{M.~Vargas-Maga\~na}\orcidlink{0000-0003-3841-1836},}
\author[22]{{B.~A.~Weaver}}

%% file: HOD_author_list.affiliations.tex
% Author list file generated with: mkauthlist 1.3.0+14.gcc6daf1.dirty 
%% Affiliations file. load \usepackage{hanging}. Use \input to call it after \appendix

\section{Author Affiliations}
\label{sec:affiliations}

\noindent \hangindent=.5cm $^{1}${Institute of Cosmology and Gravitation, University of Portsmouth, Dennis Sciama Building, Portsmouth, PO1 3FX, UK}

\noindent \hangindent=.5cm $^{2}${Department of Physics and Astronomy, University of Waterloo, 200 University Ave W, Waterloo, ON N2L 3G1, Canada}

\noindent \hangindent=.5cm $^{3}${Perimeter Institute for Theoretical Physics, 31 Caroline St. North, Waterloo, ON N2L 2Y5, Canada}

\noindent \hangindent=.5cm $^{4}${Waterloo Centre for Astrophysics, University of Waterloo, 200 University Ave W, Waterloo, ON N2L 3G1, Canada}

\noindent \hangindent=.5cm $^{5}${IRFU, CEA, Universit\'{e} Paris-Saclay, F-91191 Gif-sur-Yvette, France}

\noindent \hangindent=.5cm $^{6}${Sorbonne Universit\'{e}, CNRS/IN2P3, Laboratoire de Physique Nucl\'{e}aire et de Hautes Energies (LPNHE), FR-75005 Paris, France}

\noindent \hangindent=.5cm $^{7}${Departament de F\'{\i}sica Qu\`{a}ntica i Astrof\'{\i}sica, Universitat de Barcelona, Mart\'{\i} i Franqu\`{e}s 1, E08028 Barcelona, Spain}

\noindent \hangindent=.5cm $^{8}${Institut d'Estudis Espacials de Catalunya (IEEC), 08034 Barcelona, Spain}

\noindent \hangindent=.5cm $^{9}${Institut de Ci\`encies del Cosmos (ICCUB), Universitat de Barcelona (UB), c. Mart\'i i Franqu\`es, 1, 08028 Barcelona, Spain.}

\noindent \hangindent=.5cm $^{10}${University of Michigan, Ann Arbor, MI 48109, USA}

\noindent \hangindent=.5cm $^{11}${Laboratoire de Physique Subatomique et de Cosmologie, 53 Avenue des Martyrs, 38000 Grenoble, France}

\noindent \hangindent=.5cm $^{12}${Center for Astrophysics $|$ Harvard \& Smithsonian, 60 Garden Street, Cambridge, MA 02138, USA}

\noindent \hangindent=.5cm $^{13}${Department of Physics, The University of Texas at Dallas, Richardson, TX 75080, USA}

\noindent \hangindent=.5cm $^{14}${NASA Einstein Fellow}

\noindent \hangindent=.5cm $^{15}${Ecole Polytechnique F\'{e}d\'{e}rale de Lausanne, CH-1015 Lausanne, Switzerland}

\noindent \hangindent=.5cm $^{16}${Physics Dept., Boston University, 590 Commonwealth Avenue, Boston, MA 02215, USA}

\noindent \hangindent=.5cm $^{17}${Dipartimento di Fisica ``Aldo Pontremoli'', Universit\`a degli Studi di Milano, Via Celoria 16, I-20133 Milano, Italy}

\noindent \hangindent=.5cm $^{18}${Department of Physics \& Astronomy, University College London, Gower Street, London, WC1E 6BT, UK}

\noindent \hangindent=.5cm $^{19}${Lawrence Berkeley National Laboratory, 1 Cyclotron Road, Berkeley, CA 94720, USA}

\noindent \hangindent=.5cm $^{20}${Institute for Computational Cosmology, Department of Physics, Durham University, South Road, Durham DH1 3LE, UK}

\noindent \hangindent=.5cm $^{21}${Instituto de F\'{\i}sica, Universidad Nacional Aut\'{o}noma de M\'{e}xico,  Cd. de M\'{e}xico  C.P. 04510,  M\'{e}xico}

\noindent \hangindent=.5cm $^{22}${NSF NOIRLab, 950 N. Cherry Ave., Tucson, AZ 85719, USA}

\noindent \hangindent=.5cm $^{23}${Kavli Institute for Particle Astrophysics and Cosmology, Stanford University, Menlo Park, CA 94305, USA}

\noindent \hangindent=.5cm $^{24}${SLAC National Accelerator Laboratory, Menlo Park, CA 94305, USA}

\noindent \hangindent=.5cm $^{25}${Institut de F\'{i}sica d’Altes Energies (IFAE), The Barcelona Institute of Science and Technology, Campus UAB, 08193 Bellaterra Barcelona, Spain}

\noindent \hangindent=.5cm $^{26}${Departamento de F\'isica, Universidad de los Andes, Cra. 1 No. 18A-10, Edificio Ip, CP 111711, Bogot\'a, Colombia}

\noindent \hangindent=.5cm $^{27}${Observatorio Astron\'omico, Universidad de los Andes, Cra. 1 No. 18A-10, Edificio H, CP 111711 Bogot\'a, Colombia}

\noindent \hangindent=.5cm $^{28}${Institute of Space Sciences, ICE-CSIC, Campus UAB, Carrer de Can Magrans s/n, 08913 Bellaterra, Barcelona, Spain}

\noindent \hangindent=.5cm $^{29}${Fermi National Accelerator Laboratory, PO Box 500, Batavia, IL 60510, USA}

\noindent \hangindent=.5cm $^{30}${Department of Astrophysical Sciences, Princeton University, Princeton NJ 08544, USA}

\noindent \hangindent=.5cm $^{31}${Center for Cosmology and AstroParticle Physics, The Ohio State University, 191 West Woodruff Avenue, Columbus, OH 43210, USA}

\noindent \hangindent=.5cm $^{32}${Department of Physics, The Ohio State University, 191 West Woodruff Avenue, Columbus, OH 43210, USA}

\noindent \hangindent=.5cm $^{33}${The Ohio State University, Columbus, 43210 OH, USA}

\noindent \hangindent=.5cm $^{34}${School of Mathematics and Physics, University of Queensland, 4072, Australia}

\noindent \hangindent=.5cm $^{35}${Instituci\'{o} Catalana de Recerca i Estudis Avan\c{c}ats, Passeig de Llu\'{\i}s Companys, 23, 08010 Barcelona, Spain}

\noindent \hangindent=.5cm $^{36}${Department of Physics and Astronomy, Siena College, 515 Loudon Road, Loudonville, NY 12211, USA}

\noindent \hangindent=.5cm $^{37}${Departament de F\'isica, EEBE, Universitat Polit\`ecnica de Catalunya, c/Eduard Maristany 10, 08930 Barcelona, Spain}

\noindent \hangindent=.5cm $^{38}${Department of Physics and Astronomy, Sejong University, Seoul, 143-747, Korea}

\noindent \hangindent=.5cm $^{39}${CIEMAT, Avenida Complutense 40, E-28040 Madrid, Spain}

\noindent \hangindent=.5cm $^{40}${Department of Physics, University of Michigan, Ann Arbor, MI 48109, USA}

\noindent \hangindent=.5cm $^{41}${Department of Physics \& Astronomy, Ohio University, Athens, OH 45701, USA}